\documentclass{iopart}

\usepackage{iopams}

\usepackage{psfrag}
\usepackage{graphicx}
\usepackage{dcolumn}
\usepackage{bm}
\usepackage{epsfig}
\usepackage{yfonts}
\usepackage{version}
\usepackage[utf8]{inputenc}

\usepackage{color, soul}
\usepackage[usenames,dvipsnames]{xcolor}
\usepackage{ulem}  
\newcommand{\eqref}[1]{(\ref{#1})}
\newcommand{\onlinecite}[1]{\cite{#1}}

\newcommand{\bra}[1]{\langle\,{#1}\, |}
\newcommand{\ket}[1]{|\,{#1}\,\rangle}

\newcommand{\ElTransE}{\mathcal{E}}
\newcommand{\HamTot}{H}

\newcommand{\ScalePar}{\sigma}
\newcommand{\StabIndex}{\alpha}

\newcommand{\p}{\mathcal{P}}
\newcommand{\W}{W}
\newcommand{\Dmon}{\Delta_{\rm mon}}
\newcommand{\Dagg}{\Delta_{\rm agg}}
\newcommand{\Pperf}{P_{\rm ns}}
\newcommand{\Pout}{P^{\rm out}_{\StabIndex,\ScalePar}}
\newcommand{\overbar}[1]{\mkern 2mu\overline{\mkern-2mu#1\mkern-2mu}\mkern 2mu}
\newcommand{\Nseg}{\overbar{N}^{\rm seg}}
\newcommand{\Nloc}{\overbar{N}^{\rm loc}}
\newcommand{\jigsaw}{zigzag}

\excludeversion{weg}

\begin{document}

\title[One-dimensional excitonic systems with diagonal L\'{e}vy disorder]{One-dimensional excitonic systems with diagonal L\'{e}vy disorder: A detailed
study of the absorption spectra and the disorder scaling of localization length}

\author{S.~M\"obius$^1$, S.~M.~Vlaming$^1$, V.~A.~Malyshev$^{2,3}$, J.~Knoester$^2$, A.~Eisfeld$^1$}

\address{${}^1$ Max Planck Institute for the Physics of Complex Systems,
N\"othnitzer Strasse 38, D-01187 Dresden, Germany}
\address{${}^2$ Centre for Theoretical Physics and Zernike Institute
for Advanced Materials, University of Groningen, Nijenborgh 4, 9747
AG Groningen, The Netherlands}
\address{${}^3$ Saint Petersburg University, 198504 Saint-Petersburg, Russia}

\date{\today}

\begin{abstract}

We present a detailed study of the effects of L\'evy $\alpha$-stable disorder 
distributions on the optical and localization properties of excitonic systems. These 
distributions are a generalization of commonly studied Gaussian 
randomness ($\alpha = 2$). However, the more general case ($\alpha < 2$) includes 
also heavy-tailed behavior with a divergent second moment. These types of distributions give 
rise to novel effects, such as exchange broadening of the absorption spectra and 
anomalous localization of the excitonic states, which
has been found in various excitonic systems. We provide a thorough examination 
of the localization behavior and the absorption spectra of excitons in chains with L\'evy $\StabIndex$-stable diagonal disorder.
Several regimes are considered in detail: 
(i) weak disorder and small systems, where finite size effects become significant; 
(ii) intermediate disorder and/or larger systems for which exciton states around 
the exciton band edge spread over a number of monomers, and (iii) strong
disorder where the exciton wave functions resemble the individual molecular 
states. We propose analytical approaches to describe the disorder scaling of 
relevant quantities for these three regimes of localization, which match
excellently with numerical results. We also show the appearance of short localization 
segments caused by outliers (monomers having transition energies larger 
than the exciton band width). Such states give rise to an additional structure 
in the absorption spectrum, density of states, and in the nontrivial disorder 
scaling of the absorption band width. Finally, we provide an analysis of the effect of
truncating the tails of the disorder distributions, such that unrealistically large 
fluctuations of the monomer transition energies are excluded. It is shown that upon truncation
our analysis of the three localization regimes mentioned above still holds, with the exception that the 
overall absorption width is narrowed due to the scarcity of short segments and 
their corresponding absorption at the high energy side of the absorption band.

\end{abstract}

\maketitle

\section{Introduction}

Organic supramolecular systems, such as molecular aggregates,~\cite{Jelley1936, Scheibe1936, Kobayashi2012, Wurthner2011, Saikin2013,Mueller2013, Eisele2009, Eisele2012, Sperling2010, Abramavicius2012, Malyukin2012, Yamagata2012}
photosynthetic complexes,~\cite{Brixner2005, Engel2007, Collini2010,Aspuru-Guzik_Chlorosomes, Jendrny2012, barford2007}
and organic photovoltaic systems~\cite{Tischler2005,Bredas2009, Akselrod2012}
, are currently of great interest due to their potential applications in nano-photonic and optoelectronic devices. 
In all these objects it is of fundamental importance to obtain a detailed understanding of the nature of excitations and the influence of environmental degrees of freedom on optical response and transport properties.

The characteristic property of the above mentioned assemblies is that they are composed of closely spaced individual monomers which can exchange excitation energy via transfer dipole-dipole interactions. 
As a result, the eigenstates of the interacting monomers are coherently shared by many units.
Besides molecular systems, such a behavior can also be found in a variety of other physical objects, such as, for example, 
assemblies of ultra-cold Rydberg atoms ~\cite{AtEiRo08_045030_,RoHeTo04_042703_,MoeWueAt11_184011_,LiTaGa05_173001_}, 
quantum dots ~\cite{UnMuLi05_137404_} or superconducting circuits ~\cite{MoReEi12_105013_,HiRePl06_1427_}. 
From now on, we will denote such systems as quantum aggregates and refer to the collective optical excitations as Frenkel excitons.

The interaction of the monomer units with the often disordered (and sometimes fluctuating) environment leads to local variations in the physical parameters describing each monomer. This leads to a localization of the Frenkel excitons, which will no longer extend throughout the entire system but will be localized in smaller subparts of the system \cite{Anderson1958, ABRAHAMS1979, Kramer1993}.
The typical spatial extent of such an excitation in a disordered environment is 
referred to as the localization length ~\cite{Anderson1958, ABRAHAMS1979, Thouless1974}.

In modeling the behavior of the collective excitations in a disordered environment, 
one  often considers model parameters as 
stochastic quantities described by some probability distribution. When 
the environment changes slowly with respect to the processes 
within the system of interest, one can approximate 
the disorder induced by interactions with the host material as being static in time. 
In this work, we will 
consider the situation where static disorder influences primarily the transition energies of the monomers and 
does 
not lead to appreciable changes of the interaction between the monomers, and in particular does not lead to 
geometrical 
changes.~\cite{FiSeEn08_12858_,SeEi12_024109_,ScSeLi13_792_}
This approach has been succesfully used to model, for example, 
the absorptive and emissive properties 
of various molecular aggregates at low 
temperatures.~\cite{Malyshev2003b, BeMaKn03_217401_, Heijs2005a, Fidder1991, Augulis2010, Vlaming2009, Stradomska2010, Boukahil1990}
Note that we do not consider the coupling of a dipole transition to the monomeric vibrations (see e.g.~\cite{ScFi84_269_,MPaMa13_064703_}) or inter-monomer dynamics \cite{Halperin1966, Lifshits1968, Knapp1984}.
The localization of collective excitations has profound effects on experimentally accessible quantities related to the system under consideration, such as absorption, emission and excitation dynamics.
The localized states tend to form around the bare exciton band edges, and in particular 
will also form 
a tail in the density of states, the so called Lifshits tail ~\cite{Halperin1966, Lifshits1968}. The states in the Lifshits 
tail play an important role in determining the optical and energy transport properties of the 
material.

Disorder in the transition energies of the monomers leads to a monomer absorption spectrum that follows the disorder probability distribution. 
The most common choices for the randomness of the monomer transition energies  are box-like or Gaussian distributions.
For such disorder, the collective nature of the excitations in interacting aggregates of monomers is reflected in a reduced width of the exciton absorption peak with respect to the monomer peak, an effect which is referred to as exchange narrowing~\cite{Knapp1984, Malyshev1999, Knoester1993}.

However, it has been suggested that for certain environments more general distributions are needed \cite{BaSiZu00_5853_,BaNaVa03_075502_,KhZu02_5107_}. Such distributions can lead to new effects~\cite{Stoneham1969, Vlaming2009a, Eisfeld2010, Werpachowska2012}. Indeed, when one generalizes to the wider class of L\'evy $\alpha$-stable distributions,~\cite{Levy24, Feller66} the behavior of the exciton states can be strikingly different. For example, the well-known effect of exchange narrowing need not occur, but for sufficiently heavy-tailed distributions a broadening effect can occur instead.~\cite{Eisfeld2010}

In this paper, we consider the symmetric L\'evy $\alpha$-stable distributions as a generalization of the aforementioned Gaussian distributions for the molecular transition energies, which includes Gaussian as well as Lorentzian distributions as special cases. Such symmetric L\'evy stable distributions are determined by two parameters:  the index of stability $\alpha$ which fixes the general shape and the behavior for large arguments, and a scale parameter $\sigma$ which determines the width of the distribution. The sum of two random variables which belong to distributions with the same $\alpha$ will also belong to a distribution with the same $\alpha$. We will refer to these distributions as $\alpha$-stable distributions, L\'evy-stable distributions or simply as stable distributions.

For $\alpha=2$ and $\alpha=1$, these distributions coincide with Gaussian and Lorentzian distributions respectively. For values of $\alpha<2$, these distributions will have more weight in the tails than Gaussians. As a result, transition energies occuring in the tail are increasingly likely, and  have a significant impact on the localization behavior of the exciton states and, in turn, also on the optical properties of the system.

In a first study~\cite{Eisfeld2010} several of us have already investigated basic consequences of the heavy-tailed $\alpha$-stable distributions. In particular it was shown that for $\alpha>1$  exchange narrowing occurs, while for $\alpha<1$ the spectrum broadens upon aggregation. In addition, indications were found that outliers, defined as transition energies which are larger than the effective interaction between the monomers, can lead to unexpected effects in the optical properties. Here we will give a much more systematic discussion of the optical properties, and the closely related localization behavior, of one-dimensional aggregates with uncorrelated $\alpha$-stable disorder. 
Depending on the type of system one considers, their finite size of such systems may bear relevance to the optical properties. Therefore, we discuss exciton localization and its optical consequences for a much wider range of parameters than was done in Ref~\cite{Eisfeld2010}; that is, we scan over the parameter space provided by the stability index $\alpha$, the disorder strength $\sigma$ and the system size $N$. 

The system is shown to provide a rich interplay between the various relevant length scales. The first length scale is simply provided by the standard localization length as induced by relatively small variations of the transition energies. The second length scale is the segment length, that is, the typical distance between neighboring outliers, which provides another mechanism of exciton localization. Finally, the finite size of the system is the third length scale. Depending on the relative magnitudes of these three length scales, various types of behavior may be identified, and  we will show that the introduction of the finite system size as an additional length scale in the problem can be neatly incorporated into the theoretical description. 

Generally, three regimes can be identified. Firstly, for small disorder strengths we essentially have fully delocalized exciton states, and we can provide analytical expressions for the absorption lineshape. Secondly, for increasing amounts of disorder, the localization length drops below the system size. In this intermediate disorder regime, we will show that we can make scaling arguments for the dependence of localization length and the absorption peak width on the amount and type of disorder. Finally, for very strong disorder we obtain strongly localized states that closely resemble the molecular excited states. The segmentation effect produced by outliers, as mentioned earlier, can provide additional features in the localization behavior and absorption spectra. In addition, we can predict for which parameters each regime is applicable, allowing for the construction of 'phase diagrams' for localization behavior.

As mentioned before, L\'evy distributions (for $\alpha \neq 2$) possess different characteristics from Gaussians, including (but not limited to) an increased weight in the tails. However, in reality such environmentally induced fluctuations can not have arbitrarily large values. Therefore, in Section \ref{sec:truncated} we provide a study where we consider L\'evy distributions with truncated tails.
We demonstrate that the same three regimes apply, for very similar parameter values. A notable caveat is that there is not a proper strong localization limit, since the width of the disorder distribution remains bounded as a result of the truncation. The removal of the tails does lead to a strong reduction in outliers and segmentation, and a corresponding lack of segmentation effects. In the absorption spectrum, this removes the absorption caused by relatively short segments that would otherwise occur at the high energy side of the absorption peak. Nevertheless, this does not affect the scaling behavior of the peak width in the scaling regime.

The paper is organized as follows:\\
In Sec.~\ref{sec:model}, we introduce the basic components of our theory. We introduce the model Hamiltonian and some basic aspects of the eigenstates in Sec. \ref{sec:model_hamiltonian}. Then, the $\alpha$-stable disorder distributions are introduced in Sec.  \ref{sec:distributions}, while we also discuss some relevant mathematical properties that will be of use in the remainder of the paper. In Sec.~\ref{sec:def_absorb}, we discuss how to calculate the linear absorption spectrum, a quantity that can straightforwardly be compared to experimental measurements, and which reflects the delocalized nature of the excitations.

In Sec.~\ref{sec:analytical}, we provide analytical estimates of how certain quantities depend on the type of disorder distribution used: in particular, we concern ourselves with the mean localization length, the distribution of localization lengths, the absorption spectra and their peak widths. First, as a useful reference case, we revisit the eigenstates and eigenenergies of the disorder-free Frenkel exciton Hamiltonian in Sec.~\ref{Sec:hom}. We use these eigenstates as a basis set in the general disordered case, as is done in Sec.~\ref{sec:analytical_estimates}, where we discuss some analytical results. We can identify three regimes: the weak disorder limit, where states are strongly delocalized,  is treated in Sec.~\ref{sec:weak_disorder}, while we derive scaling relations for the dependence of the localization length on disorder  in the intermediate disorder regime in Sec.~\ref{sec:scaling_intermediate_disorder}. In the same section, we also discuss the third regime, where large disorder leads to 
strong localization effects. Sec.~\ref{sec:outlier} contains a discussion of the frequency of outliers and the typical length scales of the corresponding localization segments. Finally, we show in Sec.~\ref{sec:summary} that we can generally estimate which of the three regimes is applicable for a given combination of parameters, which can be summarized in a ``phase diagram''.

Sec.~\ref{sec:numerics} compares numerical results with the analysis presented in Sec.~\ref{sec:analytical}. First, we discuss the mean localization length and the localization length distributions in Sec.~\ref{sec:num_localization_length}, where the numerics generally agree well with the theoretical predictions in all three aforementioned regimes. In Sec.~\ref{sec:fwhmscaling},  we analyze the dependence of the absorption peak width on disorder, and show again a good agreement with our previous analysis and the occurrence of three regimes. However, a few additional features appear in the numerical results: their origin is discussed in \ref{sec:scal_50}, where we also provide a discussion of the localization patterns of the wave functions, the so-called hidden structure. Finally, we discuss the effect of truncating the tails of the $\alpha$-stable distributions in Sec.~\ref{sec:truncated}, which physically corresponds to limiting the maximum possible deviations of the transition energy. This alters some of 
the effects related to the outlier induced segmentation.

We present a final overview in Sec. \ref{sec:conclusions}.

\section{The model}
\label{sec:model}

In this section, we introduce the basic ingredients of our model, which is essentially a Frenkel exciton Hamiltonian in one dimension. We  discuss the $\alpha$-stable distributions, which we will use as the disorder distributions for the site energies and provide their relevant mathematical properties. Finally, we will describe how to calculate linear absorption spectra in excitonic systems, which is the prime experimentally observable quantity that can be used to probe these systems and study the effect of disorder.
 
\subsection{The Hamiltonian}
\label{sec:model_hamiltonian}

We consider  a one-dimensional tight-binding Hamiltonian consisting of $N$ sites. In the context of molecular aggregates, each of the sites corresponds to an individual (two-level) molecule $n$. We denote the state where molecule $n$ is in the excited state and all others are in their ground states by $\ket{\pi_{n}}$. The transfer interactions between sites $n$ and $m$ are denoted by $V_{nm}$ where $n\neq m$.  
Taking the $\ket{\pi_n}$ states as a basis for the one-excitation manifold, the Hamiltonian  takes the form
\begin{equation}
\label{Ham_Ex}
\HamTot=\sum_{n=1}^N (\ElTransE+D_n) \ket{\pi_n}\bra{\pi_n}+ \sum_{nm}^{} V_{nm}\ket{\pi_n}\bra{\pi_{m}}
\end{equation} 
where $\ElTransE$ is the mean  transition energy of the monomers 
  and $D_n$ is a random, time independent offset energy.
The presence of the latter term is e.g.~ the result of interactions of the molecules with slowly fluctuating environmental factors, and is referred to as static disorder.

In many applications of the model, the interaction can be of a long-range nature. In the context of molecular aggregates, for example, the molecules interact through their transition dipoles, e.g.\ $V_{nm}\sim 1/R_{nm}^3$ where $R_{nm}$ is the distance between monomers $n$ and $m$. However,  we have found that the qualitative physical behavior of collective excitations for such longer-range interactions is very similar to the situation for nearest-neighbor interactions.  
For excitation transport, this  similarity between long-range and nearest-neighbor interaction has been reported in Ref.~\cite{Vlaming2013}.

We limit ourselves in this work to nearest-neighbor interactions, since this allows for simple analytical results.
Then, the Hamiltonian (\ref{Ham_Ex}) becomes
\begin{eqnarray}
\label{Ham_ExNN}
 \HamTot = \sum_{n=1}^N (\ElTransE+D_n) \ket{\pi_n}\bra{\pi_n} 
  + V \sum_{n=1}^N\big(\ket{\pi_n}\bra{\pi_{n+1}}+\ket{\pi_{n+1}}\bra{\pi_{n}}\big)
\end{eqnarray}
where $V\equiv V_{n,n+1}$ denotes the nearest neighbor interaction.

The exciton states $\ket{\phi_j}$ are simply the eigenstates of the above Hamiltonian, and can generally be written as a coherent superposition of local excitations 
\begin{equation}
\label{eq:eigVal}
\ket{\phi_{j}}= \sum_{n=1}^N c_{nj} \ket{\pi_n}.
\end{equation}
In the case of no disorder ($D_n=0$), the exciton states will extend over all sites $n$. This is discussed in more detail in Sec.~\ref{Sec:hom}. However, when disorder is present, localization can occur, leading to exciton states that are coherently shared by only part of the chain's sites. The localization length is the typical extent of such an exciton state, i.e.\ a measure of the number of molecules contributing to an exciton state $\ket{\phi_j}$, and can be quantified in various ways. In the following, as is often done to quantify the localization length,~\cite{Fidder1991, Thouless1974, Schreiber1982} we will use the \emph{Inverse Participation Number}, defined as 
\begin{equation}
\label{eq:PN}
N^{\rm loc}_j=\left(\sum_n |c_{jn}|^4\right)^{-1}.
\end{equation}
One can straightforwardly check the correct behavior of this expression in the limits of a completely localized state ($c_{jn}=\delta_{nn_0}$) and a state shared equally by all molecules ($c_{jn}=1/\sqrt{N}$), where it gives $N^{\rm loc}_j=1$ and $N^{\rm loc}_j=N$ respectively. While the Inverse Participation Number is a reasonable measure for the localization length, one should keep in mind that there are different ways of quantifying localization and that there remains a certain arbitrariness as to what measure is used.

\subsection{The disorder distributions}\label{sec:distributions}

It is assumed that for each monomer the fluctuations $D_n$ are random variables distributed according to some probability distribution
\footnote{Usually this quantity is referred to as probability density. However, in this work we will use the terms probability distribution and probability density equivalently.}
$\p(D_n)$.
The $D_n$ are taken to be independent and identically distributed. 

\begin{figure}[tp]
\includegraphics[width=6cm]{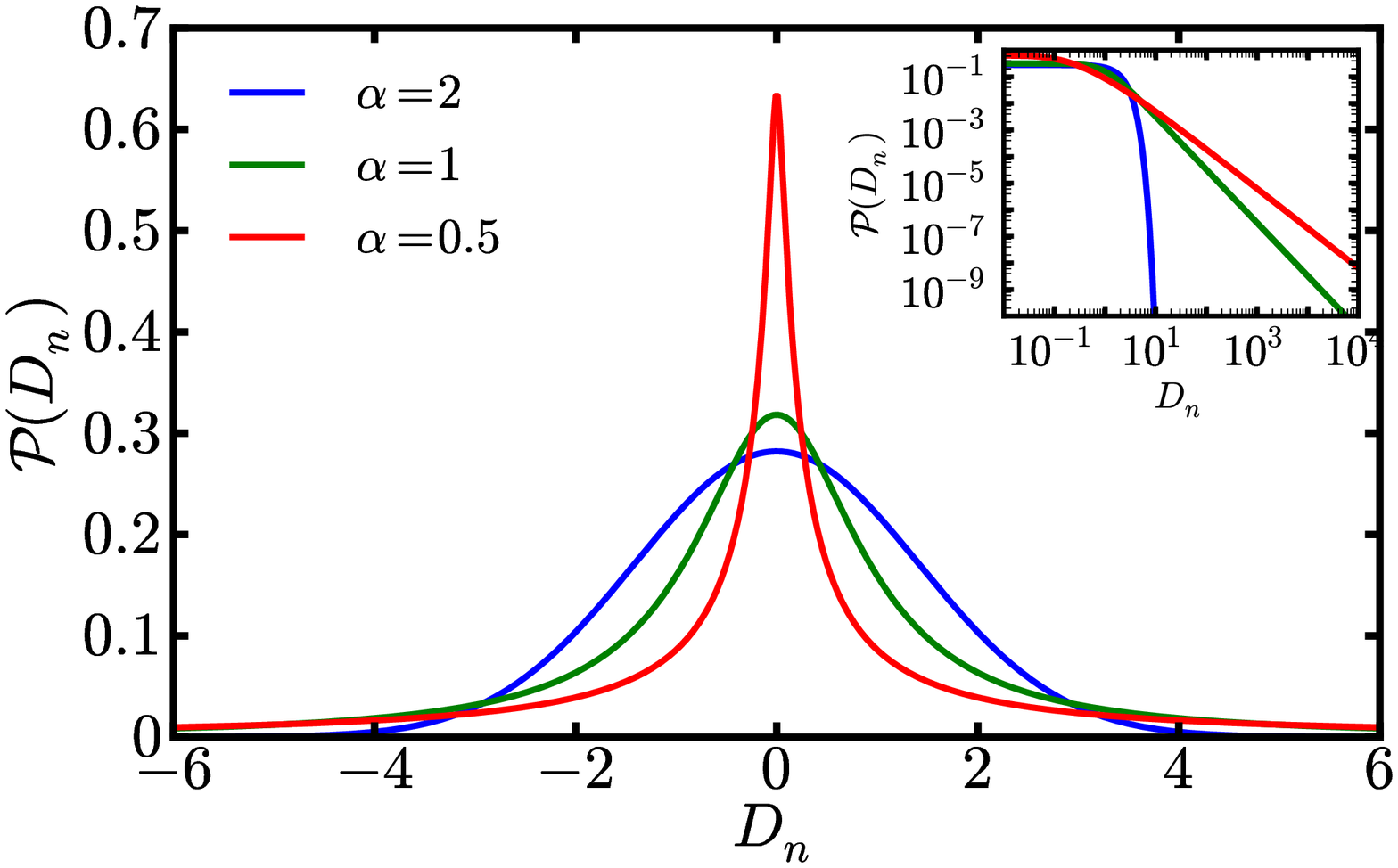} 
\includegraphics[width=6cm]{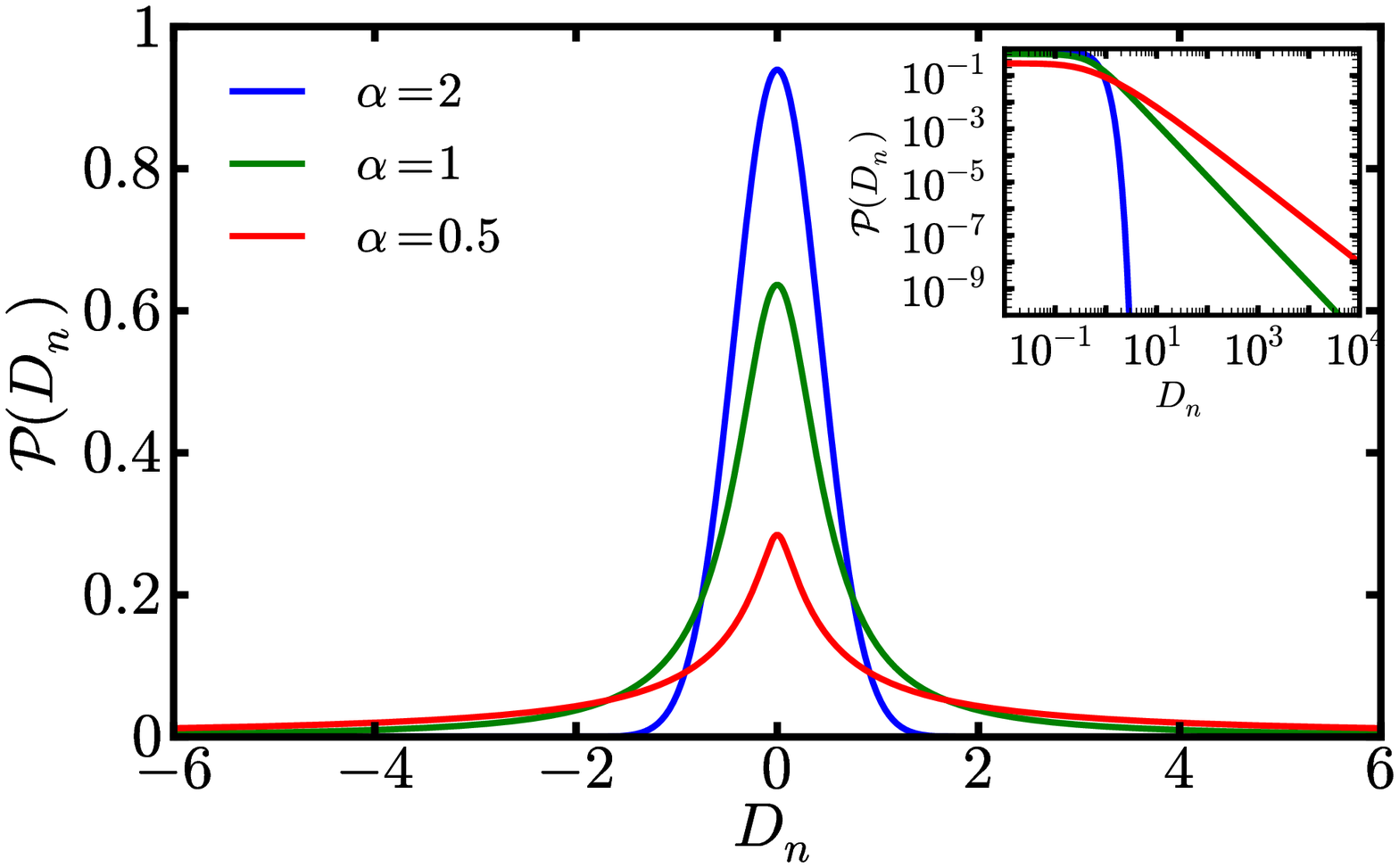}
\caption{\label{dist_examples}Probability densities for $\StabIndex$-stable distributions with $\StabIndex=2$ (blue), $1$ (green) and $0.5$ (red) shown for linear (main graph) and double logarithmic (inset) axes. Left: For all curves $\ScalePar=1$. Right: For all curves FWHM$\equiv\Delta=1$.  }
\end{figure}

In this work we will focus on  symmetric, $\alpha$-stable probability densities $\p(D_n)$ defined by
\begin{equation}
\label{fourier}
    \p_{\StabIndex,\ScalePar}(D_n) = \frac{1}{2 \pi}\int\limits_{-\infty}^{\infty} \varphi_{\StabIndex, \ScalePar}(\nu)\, e^{- \mathrm{i}\nu D_n}\,\mathrm{d}\nu 
\end{equation}
where the characteristic function $ \varphi(\nu)$ is given by
\begin{equation}
\label{def:stable_char}
\varphi_{\StabIndex,\ScalePar}(\nu)=\mathrm{e}^{-\ScalePar^{\StabIndex}|\nu|^{\StabIndex}}, \hspace{1cm}0<\StabIndex\le2, \hspace{0.2cm} \ScalePar>0
\end{equation}
with two parameters, the \textit{index of stability} $\StabIndex$ and the \textit{scale parameter} $\ScalePar$. Note that the above is a rather general class of distributions.
Examples of the distributions defined by Eq.~\eqref{fourier} and \eqref{def:stable_char} are shown in Fig.~\ref{dist_examples}.
Note that the characteristic function  completely and uniquely defines the probability distribution.

In general, the distributions defined by Eqs.~(\ref{fourier}) and (\ref{def:stable_char}) have no simple representations in terms of elementary functions \cite{Samorodnitsky1994}. 
However, several well-known distributions are special cases of the $\alpha$-stable distributions. In particular, for $\StabIndex=2$ we simply recover a Gaussian distribution with zero mean and standard deviation $\sqrt{2} \sigma$. Note that this notation differs from the usual convention where $\sigma$ is used to denote the standard deviation. 
Similarly, for $\StabIndex=1$, the probability density $\p_{1,\ScalePar}$ corresponds to a Lorentzian (Cauchy) distribution $\p_{1,\ScalePar}(D_n) = \ScalePar/\pi ({D_n}^2+\ScalePar^2)$.
In section \ref{sec:numerics} we will present numerical results for the cases of $\StabIndex = 2$, $\StabIndex = 1$ and $\StabIndex = 0.5$, as typical representations of $\StabIndex$-stable distributions. For the latter value of $\StabIndex = 0.5$, there also exists an analytic expression of $\p_{\StabIndex,\ScalePar}$ in terms of Fresnel integrals and trigonometric functions \cite{Hopcraft1999}.

For $\StabIndex <2$ the tails of the distributions behave like $\p_{\StabIndex,\ScalePar}(D_n)\sim 1/D_n^{\StabIndex+1}$ for large values of $D_n$ (cf.\ insets of Fig.~\ref{dist_examples}), i.e.\ they have heavy tails.
That implies that for $\StabIndex<2$ the second moment 
diverges, and the standard deviation can thus not be used as a measure of the width of the distribution. 
Therefore, we will instead use the full width at half maximum (FWHM), denoted by $\Delta$, to characterize the width [One could also use a different quantity as a measurement of the width of a probability distribution, e.g.\ for L\'evy-stable distributions the scale parameter $\ScalePar$. We choose the FWHM since it is a property of the probability distribution that can be easily extracted. Furthermore we will also use the FWHM to measure the width of the absorption spectra (see Sec.~\ref{sec:def_absorb}).].

For $\alpha=2$ one obtains $\Delta$=$4\, \sqrt{\ln 2}\, \sigma$, for $\alpha=1$ one has  $\Delta$=$2\, \ScalePar$ and for $\StabIndex = 0.5$ one finds $\Delta\approx0.4476\ScalePar$. 
In the right panel of Fig.~\ref{dist_examples} the probability densities for $\StabIndex = 2$, $\StabIndex = 1$ and $\StabIndex = 0.5$ with a fixed $\Delta$ = 1 are shown as representative examples.

The following properties of the distributions Eq.\ (\ref{def:stable_char}) will be of importance in the ensuing analysis (for details, see e.g.\ Refs~\cite{Samorodnitsky1994}):

\begin{itemize}
\item[1)] {\it The scaling parameter $\ScalePar$ is a measure of the width of the distribution:}\\
Let $a$ be a non-zero, real constant. For $D_n$ belonging to a stable distibution with scale parameter $\sigma$, one finds that $(a\,D_n)$ belongs to the same stable distribution, but with a scale parameter of $|a|\sigma$.
 Equivalently, it follows from the definition Eq.~(\ref{fourier}) and  Eq.~(\ref{def:stable_char}) that
\begin{equation}
\label{dichte_trafo}
\p_{\StabIndex, |a| \ScalePar}(D_n)=\frac{1}{|a|}\p_{\StabIndex,\ScalePar}({D_n}/{ |a|}).
\end{equation}

\item[2)] {\it The sum of random variables distributed with the same $\StabIndex$ is distributed according to the same $\StabIndex$, but a rescaled width $\sigma$:}\\
\label{item_summ_alpha_stab}
Let $D_1$ and $D_2$ be independent random variables distributed according to $\p_{\StabIndex,  \ScalePar}(D_n)$. Then the sum $Y=D_1+D_2$ is distributed according to an $\alpha$-stable distribution with renormalized width $\sigma_Y=(\ScalePar_1^{\StabIndex}+\ScalePar_2^{\StabIndex})^{1/{\StabIndex}}.$
This property will be of special importance in what follows.
Note that if $\ScalePar_n\equiv \ScalePar$ for all $n$, then the average $Y=\frac{1}{N}(D_1+\cdots+D_N)$ follows a distribution with $\ScalePar_Y=N^{\frac{1}{\StabIndex}-1}\ScalePar$.
 
\item[3)]  {\it Generalized central limit theorem (CLT): }\\
L\'evy-stable distributions with $\alpha < 2$ do not have a finite variance and therefore do not obey the standard CLT. However there exists a similar theorem for heavy-tailed distributions \cite{Voit2005, Kolmogorov1968}. In short, it states that if $X_1, X_2,\dots, X_n$ are independent, identically distributed random numbers with a probability distribution that decays like $\propto |X_i|^{-(\alpha + 1)}$ for $|X_i| \gg 1$ with $0<\alpha<2$, then the probability distribution $P_n$ of the sample mean $Y_n = {\sum^n_{i=1} X_i}/{n}$ will converge to an $\StabIndex$-stable distribution,
$
 \lim\limits_{n\rightarrow\infty}  P_n(Y_n) = \p_{\alpha, \sigma}(Y)
$
, with $Y = \lim\limits_{n\rightarrow\infty} Y_n$.

\end{itemize}

\subsection{The absorption spectrum}\label{sec:def_absorb}

An interesting property of molecular aggregates is the fact that they exhibit optical properties that are markedly different from those of the individual molecules they consist of. The collective nature of the relevant excitations leads to superradiant absorption and emission, a shift in the peak position and a change in the line-shape~\cite{Fidder1991, Schreiber1982}.
In this subsection, we will discuss how to evaluate the linear absorption spectra of an aggregate, and 
provide analytical results and estimates for some special conditions.

Let $\ket{\phi_{j}}$ be an eigenstate (Eq.~\eqref{eq:eigVal}) of the Hamiltonian (\ref{Ham_Ex}) with energy $E_j$.
Then, the orientationally averaged strength for absorption from the ground state $\ket{g} = \ket{g_1}\otimes\cdots\otimes\ket{g_N}$  
to the excited eigenstate $\ket{\phi_{j}}$ of an aggregate of length $N$ is
\begin{equation}
\label{eq:A_j}
A_j=\Big|\bra{g}\sum_n \hat{\vec{\mu}}_n \ket{\phi_{j}}\Big|^2= \Big|\sum_n c_{nj} \vec{\mu}_n\Big|^2.
\end{equation}
Here, $\hat{\vec \mu}_n = \vec{\mu}_n(\ket{g_n}\bra{e_n} + \ket{e_n}\bra{g_n})$  is the transition dipole operator of monomer $n$, where $\ket{g_n}$ and $\ket{e_n}$ denote the (monomer) ground and excited state, respectively.

In this work we will focus on the basic effects of the disorder in the transition energies and not on geometrical effects. Therefore, to keep the discussions as simple as possible, all transition dipole
moments of the monomers are chosen to be parallel and we set $|\vec\mu_n|=1$.
The (frequency dependent) absorption strength of a single aggregate, i.e.\ a single realization of disorder, is then given by
\begin{eqnarray}
 \label{eq:A(E)}
 A(E)&=&\sum_{j=1}^N A_j \delta(E-E_j) \nonumber \\
 &=&\sum_{j=1}^N\Big|\sum_n c_{nj}\Big|^2 \delta(E-E_j)
\end{eqnarray}
Note that $A_j$ and $E_j$ depend on the length $N$ of the aggregate and the specific disorder realization.

The total absorption spectrum of an ensemble of aggregates is obtained by averaging over the disorder,
\begin{equation}
\label{eq:A_aver}
\mathcal{A}(E)=\left< \sum_{j=1}^N A_j \delta(E-E_j)\right>
\end{equation}
where $\langle \cdots \rangle = \int\prod_n \p(D_n) \cdots$ denotes the disorder average.
Here it is assumed that the dynamic broadening of the individual exciton transitions is negligible, which holds at low temperature~\cite{Heijs2005a}.

Note that the absorption spectrum of $N$ uncoupled monomers is simply proportional to the bare disorder probability density, $\mathcal{A}(E) = N\;\p_{\StabIndex,\ScalePar}(D_n)$.

As in the case of the probability distributions $\p_{\StabIndex, \ScalePar}$, we will use the full width at half maximum (FWHM) as a measure for the width of the absorption spectrum. 
The width of the monomer absorption spectrum,i.e.~ the bare disorder distribution width, will be denoted by $\Dmon$, as has been proposed in Ref.~\cite{Werpachowska2012}, but not the scale parameter $\sigma$ used in our previous papers \cite{Eisfeld2010} and \cite{Eisfeld2012}.
The width of the absorption spectrum of an ensemble of aggregates, i.e.\ its FWHM, is denoted by $\Dagg$.
The choice of $\Dmon$ as width characterization also leads to $\Dagg/\Dmon = 1$ for an aggregate of length $N = 1$ (or similar a very weakly interacting aggregate), instead of a value that depends on $\StabIndex$, as is the case for the choice of $\ScalePar$ as width characterization.

\section{Analytical estimates}\label{sec:analytical}

In this section, we provide analytic estimates of the absorption spectrum for various cases of the ratio $\Dmon/|V|$ of the
width of the disorder distribution and the interaction between the monomers.

In Sec.~\ref{Sec:hom}, we first consider the eigenstates, eigenenergies and absorption spectrum for an ensemble of identical, disorder-free aggregates  as a useful reference case.
We can then rewrite the general disordered Hamiltonian Eq.~(\ref{Ham_ExNN}) in the basis of the homogeneous solutions, allowing us to quantify various regimes  in the disordered case. In addition, this allows for estimates of the localization length of the relevant exciton states. 

We can identify three different parameter regimes. 
First, for very weak disorder $\Dmon/|V| \ll 1$ (or equivalently very strong interaction between the monomers) and an aggregate of finite size, the exciton states fully extend over the aggregate and finite size effects dominate. Here, a fully analytical expression for the absorption line-shape can be obtained. Secondly, localization of exciton states will occur for intermediate disorder, and for a large range of disorder strengths we can derive analytical expressions for the FWHM  $\Dagg$ as a function of the ratio $\Dmon/|V|$. Finally, for very large disorder $\Dmon/|V|\gg 1$, the exciton states are strongly localized and closely resemble the molecular excited states. All these cases are analyzed in Sec.~\ref{sec:analytical_estimates}.

In Sec.~\ref{sec:outlier}, we provide more details on the frequency of outliers, which lead to a segmentation of the chain. We identify regions in parameter space where either the conventional Anderson localization  or the segmentation effect is expected to provide the dominant localization mechanism. Finally, in Sec.~\ref{sec:summary}, we quantify where the boundaries between the aforementioned three regimes are located for given parameters. This allows us to provide a phase diagram in parameter space that shows where the various regimes are applicable.

\subsection{No disorder. Eigenstates, eigenenergies and absorption strength }\label{Sec:hom}

The disorder-free chain with nearest-neighbor interactions (i.e., Hamiltonian \eqref{Ham_ExNN} with $D_n=0$) can be solved exactly, and provides a useful reference solution for later purposes \cite{Knapp1984,Amerongen2000,Fidder1991,Malyshev1995}.
For this case, the eigenstates \eqref{eq:eigVal} are simply given by
$
\ket{\phi_{j}}= \sum_{n=1}^N c_{nj} \ket{\pi_n} 
$
with coefficients 
\begin{equation}
\label{eq:coeffNN}
c_{nj}=\sqrt{\frac{2}{N+1}}\sin\Big(\frac{\pi j n}{N+1}\Big)  
\end{equation}
and corresponding eigenenergies 
\begin{equation}
\label{eq:eigE}
E_{j}=\ElTransE+ 2V \cos\Big(\frac{\pi j }{N+1}\Big).
\end{equation}

As mentioned after Eq.~\eqref{eq:A_j} we take all transition dipole moments to be parallel and of unit magnitude.
Then the absorption strength of state $\ket{\phi_{j}}$ is \cite{Knapp1984,Fidder1991,Malyshev1995}
\begin{equation}
\label{F_nu}
A_{j}=\big|\sum_{n=1}^Nc_{nj}\big|^2=\frac{1-(-1)^{j}}{N+1}\mathrm{cot}^2\frac{\pi j}{2(N+1)}
\end{equation}
We will consider the case of negative $V$ (which leads to the formation of a
J-band \cite{EiBr06_376_}, that is, the aggregate absorption spectrum is redshifted with respect to the monomer spectrum. The state with $j=1$ is the lowest state of the exciton band and carries nearly all the oscillator strength (see Eq.~\eqref{F_nu}). This state, where the molecules all absorb in phase by additively contributing to the exciton transition dipole moment Eq.~\eqref{F_nu} (i.e., $c_{nj}$ has the same sign for all molecules $n$), is often referred to as superradiant and dominates the optical absorption. 
[This state  carries for large $N$ roughly 81\% of the oscillator strength (for  $N<7$ even more than 90\%).]
From Eq.~\eqref{F_nu} it can also be seen that states with even $j$ do not carry any absorption strength at all.

\subsection{Analytic estimates with disorder}
\label{sec:analytical_estimates}
\begin{figure}[tp]
\begin{center}
\includegraphics[width=.45\linewidth]{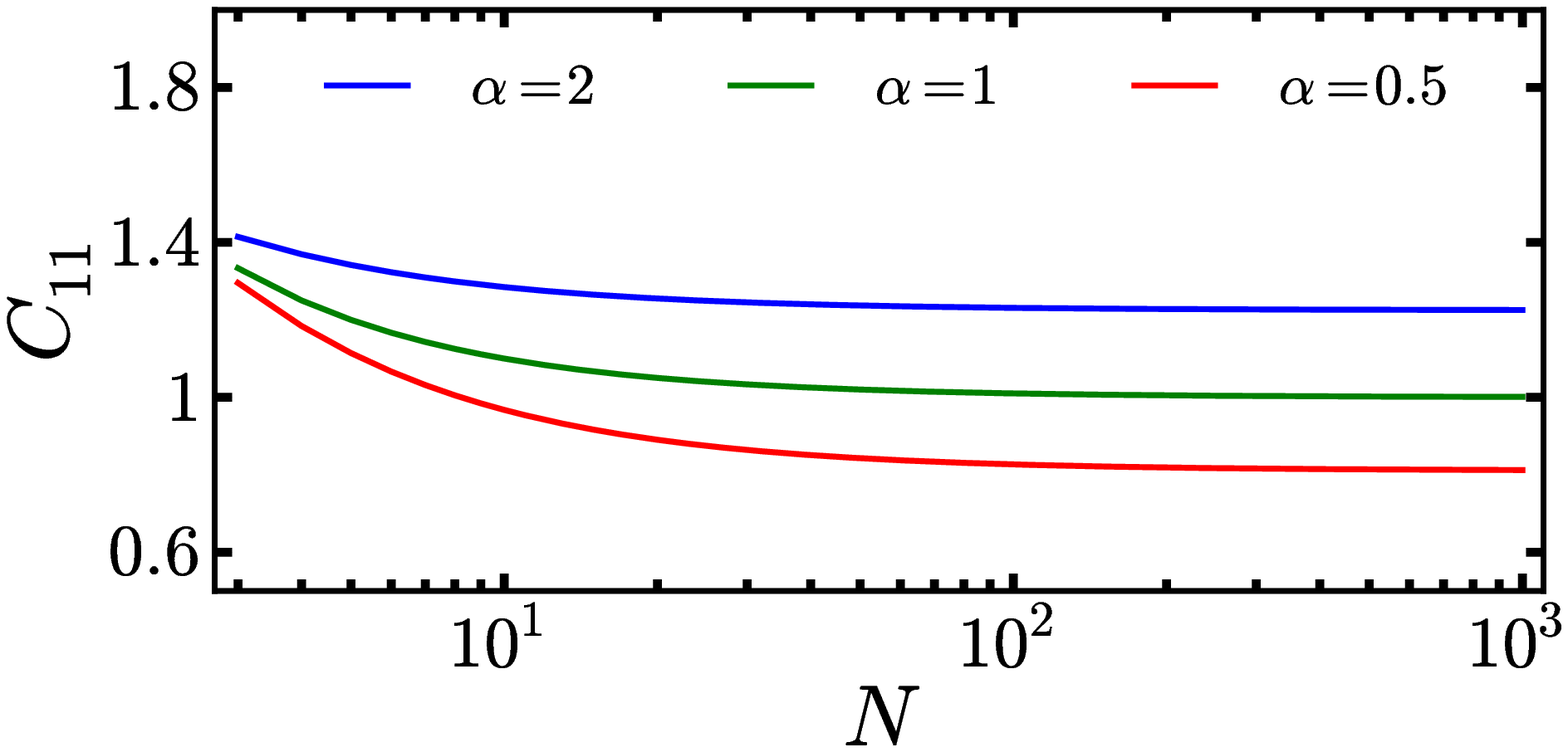}
\includegraphics[width=.45\linewidth]{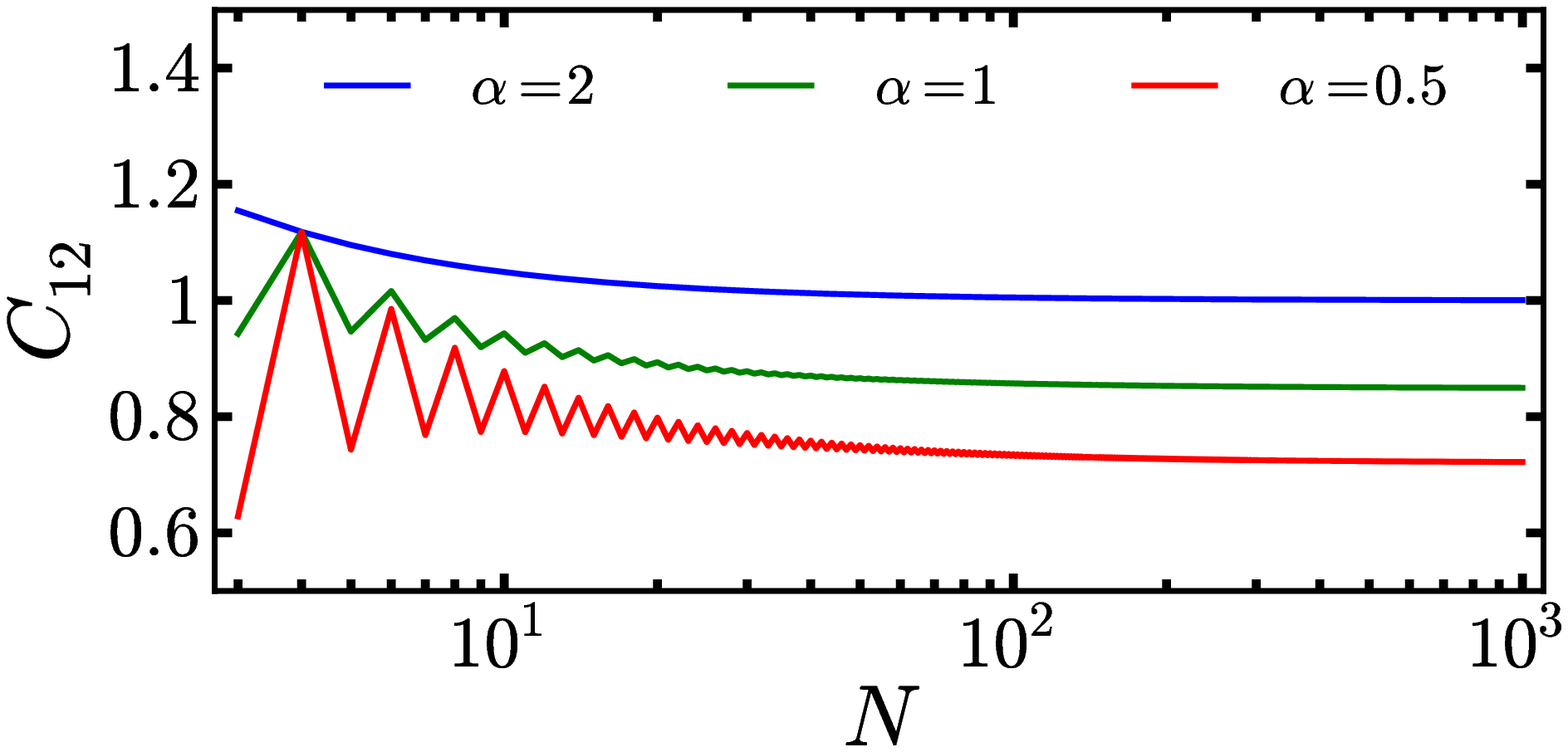}
\caption{\label{fig:c_ij_examples}
Examples of the $N$-dependence of the coefficients $C_{jj'}(N,\StabIndex)$, defined in Eq.~\eqref{eq:g_approx}, for index of stability $\StabIndex = 2$ (blue), $1$ (green) and $0.5$ (red).}
\end{center}
\end{figure}

We can use the disorder-free exciton states Eq.~\eqref{eq:coeffNN} as a basis to expand the Hamiltonian Eq.~\eqref{Ham_ExNN}, including disorder. This yields
\begin{equation}
H=\sum_{j=1}^N E_{j}\ket{\phi_{j}}\bra{\phi_{j}}+\sum_{j,j'=1}^N W_{jj'}\ket{\phi_{j}}\bra{\phi_{j'}}
\end{equation}
where
\begin{equation}
\label{Delta_nunu}
W_{jj'}=\frac{2}{N+1} \sum_{n=1}^N \ D_n\ \sin(\frac{\pi j n}{N+1})\sin(\frac{\pi j' n}{N+1}).
\end{equation}
The matrix elements $W_{jj'}$ describe the disorder-induced mixing between disorder-free eigenstates and are linear combinations of the original random variables $D_n$ -- thus, they are also random variables.
The distributions of the couplings $W_{jj'}$ are crucial to the spectral properties of the aggregate.
The diagonal terms $W_{jj}$ describe the first order disorder induced shifts of the exciton levels $E_{j}$ and the non-diagonal part $W_{jj'}$ with $j\ne j'$ 
describes disorder-induced mixing between disorder-free eigenstates (Eqs.~(\ref{eq:eigVal}) and (\ref{eq:coeffNN})).
Throughout this work we will consider the FWHM $\Delta_{jj'}$ of the distribution $W_{jj'}$ as a measure of the mixing-strength. [This is again different from our previous works \cite{Eisfeld2010, Eisfeld2012} where we have used the scale parameter $\ScalePar$ (see Sec.~\ref{sec:def_absorb}).]
For $D_n$ distributed according to the stable distribution Eq.~\eqref{dichte_trafo} with stability index $\alpha$ and scale parameter $\sigma$, one finds that the $W_{jj'}$ are distributed according to a stable distribution with the same $\alpha$ but with a different scale parameter (see Sec.~\ref{item_summ_alpha_stab}, property 2),
\begin{equation}
\label{W_jj'}
\sigma_{jj'}=\ScalePar \, g_{jj'}(N,\StabIndex )\,
\end{equation}
with
\begin{equation}\label{eq:g_jj'}
g_{jj'}(N,\StabIndex )=\frac{2}{N+1}\left(\sum_{n=1}^N \left|\sin\left(\frac{\pi j n}{N+1}\right)\sin\left(\frac{\pi j' n}{N+1}\right)\right|^{\StabIndex}\right)^{\frac{1}{\StabIndex}}.
\end{equation}
Note that $\Delta_{jj'}$ is proportional to the scale parameter $\ScalePar_{jj'}$ belonging to the distribution  of $W_{jj'}$.

Eqs.~\eqref{W_jj'} and \eqref{eq:g_jj'} are crucial in understanding the aggregate lineshape.
To obtain simple estimates, we rewrite \eqref{eq:g_jj'} in a form that makes the dependence on $N$ more apparent
\begin{equation}
\label{eq:g_approx}
g_{jj'}(N,\StabIndex)= C_{jj'}(N,\StabIndex) \cdot \frac{1}{N+1}N^{1/\StabIndex} 
\, \stackrel{\scriptscriptstyle N \gg 1}{\sim} \,  N^{\frac{1}{\StabIndex}-1}
\end{equation}

Here, one finds that the factor $C_{jj'}(N,\StabIndex)$, which in fact is defined by Eq.~\eqref{eq:g_approx}, is for all arguments $j,j',\ N,$ and $\StabIndex$ of the order of unity and shows only a weak dependence on them (for the case $(j,j') = (1,1)$ and $(1,2)$, which will be of particular interest in the following, this is shown explicitly in Fig.~\ref{fig:c_ij_examples}).
From Eq.~\eqref{eq:g_approx} one sees that the perturbations $\W_{jj'}$ have the same distribution as
the $D_j$ but with a width rescaled by roughly $N^{\frac{1}{\StabIndex}-1}$, for large $N$.

\subsubsection{Weak disorder:}
\label{sec:weak_disorder}
We first consider the case of weak disorder, i.e., $\Dmon/|V| \ll 1$, where the exciton states are strongly delocalized and the
mixing of different excitonic states $\ket{\phi_j}$ caused by the scattering elements $\W_{j \neq j'}$ is very small.
For this case, we can obtain analytical expressions for the aggregate absorption lineshape.

For weak disorder, as discussed e.g.\ in Refs.~\onlinecite{Knapp1984, Ma93_225_, DoMa04_226_, Knoester1993}, to a very good approximation, the eigenstates and the corresponding absorption strengths are still given by those of the disorder-free chain (discussed in the Sec.~\ref{Sec:hom}).
However, the eigenenergies $E_j$ are different for each disorder realization, as each eigenenergy $E_j$ is shifted by the first oder correction $W_{jj}$. The $W_{jj}$ are distributed according to Eq.~\eqref{eq:g_jj'} and the absorption spectrum is given by
\begin{equation}
\label{eq:weak_perturb}
\mathcal{A}^{\rm weak}(E)=\sum_{j=1}^{N} A_j\, \mathcal{P}_{\StabIndex,\ScalePar_j}(E-E_j)
\end{equation}
where 
$\ScalePar_j=g_{jj}(N,\StabIndex) \sigma $.
This means that the spectrum of the disordered aggregate has peaks at the same positions $E_j$ and with the same weights $A_j$ as the homogeneous linear chain. However, each peak is broadened by disorder and has the same lineshape as the bare disorder distribution, but with a rescaled 
width $\Delta_{jj}=g_{jj}(N,\StabIndex)\Dmon$. 

The aggregate absorption is dominated by absorption into the superradiant transition $j=1$, which occurs (for our case $V<0$) at the bottom of the exciton band.
Its width $\Delta_{11}$ 
is given (using  Eq.~(\ref{eq:g_jj'}) and $N\approx N+1$) by
\begin{equation}
\label{perturb_breite}
\Delta_{11}\approx C_{11} \cdot N^{\frac{1}{\StabIndex}-1} \Dmon
\end{equation}
This shows that for $\StabIndex>1$ the spectrum narrows. In particular, for a Gaussian distribution ($\StabIndex=2$), we recover the well-known $1/\sqrt{N}$
narrowing, which is the exchange narrowing effect \cite{Knapp1984, Malyshev1999}.
For $\StabIndex=1$ the width of the aggregate absorption peak remains the same, while for  $\StabIndex<1$  a broadening occurs with respect to the monomer width.
As an example, for $\StabIndex=1/2$ this broadening is proportional to the number of monomers $N$.

\begin{figure*}
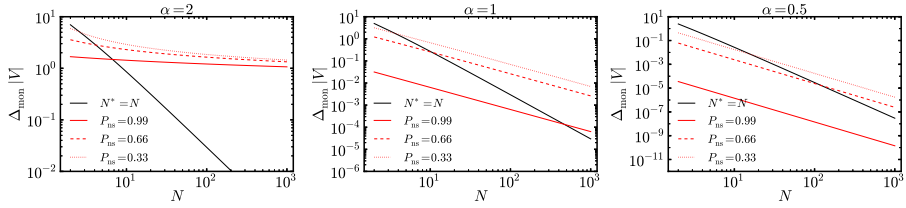

\includegraphics[width=0.3\linewidth]{{{fig_Delta_crit_alpha__2}}}
\includegraphics[width=0.3\linewidth]{{{fig_Delta_crit_alpha__1}}}
\includegraphics[width=0.3\linewidth]{{{fig_Delta_crit_alpha__0.5}}}
\caption{\label{fig:weak_valid} The black solid line shows values of $\Dmon/|V|$ for which in Eq.~\eqref{eq:weak_disorder_condition} the equal sign holds. For values of $\Dmon/|V|$ that are (far) below the black line no localization due to mixing of eigenstates is expected and  the weak disorder absorption spectrum Eq.~\eqref{eq:weak_perturb} should be valid. The red lines are isolines for which the probability to find non-segmented chains $\Pperf$ is $\Pperf = 99\%$(solid, red), $\Pperf = 66\%$ (dashed, red) and $\Pperf = 33\%$ (dotted, red). Note that the y-axes are different for each panel.
}
\end{figure*}

The above derivation of Eq.~(\ref{perturb_breite}) is only valid when the energy difference between two unperturbed exciton-states $E_j-E_{j'}$ is large compared to the disorder induced scattering $\W_{jj'}$ or equivalently $|E_{j'} - E_j| \ll \Delta_{jj'}$.
This inequality must in particular hold for the states at the band edge, for which the energetic separation is smallest. For these states one has  $E_2-E_1\approx 3 \pi^2 |V|/(N+1)^2$.
From Eqs.~\eqref{W_jj'} and \eqref{eq:g_approx} one finds for the width $\Delta_{12} \approx N^{1/\StabIndex}/(N+1)$.
Using the above expressions for $E_2 - E_1$ and $\Delta_{12}$ leads to the condition \cite{Ma93_225_, Knapp1984}
\begin{equation}
\label{eq:weak_disorder_condition}
   \frac{\Dmon}{|V|}\ll \frac{3 \pi^2}{N^{\frac{1}{\alpha}} (N+1)}, 
\end{equation}
for which the weak disorder limit and in particular Eq.~\eqref{eq:weak_perturb} is valid.
To visualize when this inequality \eqref{eq:weak_disorder_condition} holds, we show in Fig.~\ref{fig:weak_valid} for which combinations of $\Dmon/|V|$ and $N$ the inequality \eqref{eq:weak_disorder_condition} becomes an equality (black line).
For values of $\Dmon/|V|$ (far) below the black line the weak disorder limit should be valid. For $\Dmon/|V|$ above the line a mixing of states $\ket{\phi_{j}}$ and $\ket{\phi_{j'}}$ due to the coupling element $W_{jj'}$, especially $W_{12}$, will occur, leading to a localization of these states (see sec.~\ref{sec:scaling_intermediate_disorder}).
From Fig.~\ref{fig:weak_valid} on can also see that for decreasing $\StabIndex$ a smaller $\Dmon/|V|$ is needed to be in the weak disorder regime.
This can be seen for example by comparing $\Dmon/|V|$ values at the solid black line for $N = 10^2$. For $\StabIndex = 2$ (left panel of Fig.~\ref{fig:weak_valid}) on finds $\Dmon/|V|$ in the order of $10^{-2}$, for $\StabIndex = 1$ (center panel of Fig.~\ref{fig:weak_valid}) in the order of $10^{-3}$ and for $\StabIndex = 0.5$ (right panel of Fig.~\ref{fig:weak_valid}) in the order of $10^{-5}$.
This particular dependence of the weak disorder regime on $\StabIndex$ is a signature of the heavy tails of the L\'evy-stable distribution.
Only for very small widths $\Dmon/|V|$ of the bare disorder distribution are the heavy tails sufficiently insubstantial to avoid mixing of the states $\ket{\phi_j}$ and $\ket{\phi_{j'}}$.

A further effect that can change the result (\ref{eq:weak_perturb}) is segmentation caused by 'outliers'. We will discuss this concept in more detail in Sec.~\ref{sec:outlier}.
Essentially, due to the large energy difference with its neighbors, outliers only couple very weakly to the other sites. Therefore, the occurrence of outliers effectively cause a subdivision of the chain into smaller segments, each capped by two outliers.
Each of these segments will contribute peaks to the absorption spectrum, that correspond to an aggregate with the respective segment length.

\subsubsection{Scaling laws for intermediate and strong disorder:}
\label{sec:scaling_intermediate_disorder}

For increasing disorder strengths, the exciton states will localize more strongly. Above a certain $\Dmon/|V|$ the typical localization length for the optically relevant exciton states, near the lower band edge, drops below the chain length.
This occurs when the disorder strength $\Dmon$ is still much smaller than the coupling $|V|$ between the monomers, but when one is no longer in the regime where disorder induced coupling between the unperturbed eigenstates can be ignored.
In this regime we do not possess simple analytical formulas for the absorption lineshape.
However, one can derive ~\cite{Eisfeld2010} scaling laws for the FWHM $\Dagg$, as a function of the disorder strength $\Dmon$.

To obtain a simple analytic estimate for the width of the spectrum as a function of $\Dmon/|V|$ we follow the approach put forward in Ref.~\onlinecite{Ma93_225_}(see also \cite{DoMa04_226_}).
As before, we focus on the states near the bottom of the band, which are the optically dominant ones.
The mixing of various unperturbed levels by $\W_{jj'}$ leads to localized states.
At the bottom of the band, one can make an estimate of the typical localization length $N^*$ by comparing the two relevant energy scales on a localization segment, namely (i) the energy difference $E^{*}_2 - E^*_1$ between exciton states localized there, and (ii) the magnitude of the mixing $\Delta^{*}_{12}$ between the two states. Here $\Delta^*_{12}$ is the width of the distribution of the disorder induced coupling element $W^*_{12}$ for states $\ket{\phi_1}$ and $\ket{\phi_2}$ localized on $N^*$ sites. Essentially, the localization length $N^{*}$ will be such that those two energies are (approximately) equal. This argument is presented in more detail in \cite{Malyshev1995}.
We approximate the exciton level separation $E^{*}_2-E^{*}_1$ between the lowest two states on a localization segment by the level separation of the lowest two exciton states on a homogeneous chain of (effective) length $N^{*}$. Likewise, the width $\Delta_{12}^*$  of the distribution of $\W_{12}^*$ is approximated by substituting in Eqs.~\eqref{W_jj'} and \eqref{eq:g_approx} the chain length $N$ by the localization length $N^*$.

Equating $E^{*}_2-E^{*}_1\approx\Delta^{*}_{1 2}$ leads to the following equality that $N^{*}$ should obey,   
\begin{equation}
\label{eq:scaling_N*_no_approx}
\frac{3 \pi^2 |V|}{(N^{*}+1)^2} \approx \frac{(N^{*})^{\frac{1}{\StabIndex}}}{N^* + 1}\Dmon
\end{equation} 
Solving this equation for $N^{*}$ and approximate $N^* + 1\approx N^*$ leads to 
the following scaling law for the localization length,
\begin{equation}
\label{eq:scaling_N*}
N^{*}\approx \mathrm{const_1}\cdot \left(\frac{\Dmon}{|V|}\right)^{-\frac{\StabIndex}{\StabIndex+1}}
\end{equation}
The constant is given by $\mathrm{const_1}=(3 \pi^2)^{\StabIndex/(1+\StabIndex)}$.
This prefactor, besides being only a rough estimate, is somewhat arbitrary, since its value depends strongly on how one defines the width of the disorder-induced mixing distribution, i.e.~of $W^*_{12}$.
Instead of our present choice of the FWHM $\Dmon$ we could have chosen, e.g., the scale parameter or any other 'reasonable' measure for the width.  
Nevertheless, we will find in the numerical simulation, presented in section \ref{sec:num_localization_length}, that this prefactor $\mathrm{const_1}$ describes the numerical results quite well.
If one compares the right and left hand side of Eq.~\eqref{eq:scaling_N*_no_approx} with Eq.~\eqref{eq:weak_disorder_condition} one finds that they are identical except interchanging $N$ and $N^*$. This is consistent with the fact, that if the localization length $N^*$ exceeds the system size $N$, one is in the weak disorder regime with fully delocalized states.

By using the above estimate for the typical localization length of the superradiant states near the bottom of the exciton band, we can estimate the width of the absorption peak. Since the exciton states are now localized on domains with average length $N^{*}$, in Eq.~\eqref{eq:g_approx} the total number of monomers $N$ is replaced by the  (effective) number $N^*$ of coherently absorbing molecules in the respective states.
Then, one can estimate the width of the absorption band from Eq.~(\ref{perturb_breite}) with using $N^*$  instead of $N$, i.e.\ $\Dagg\approx (N^{*})^{(1-\alpha)/\alpha}\Dmon$.
With $N^{*}$ from  Eq.~(\ref{eq:scaling_N*}) one obtains
\begin{equation}
\label{eq:width_scaling}
\frac{\Dagg}{\Dmon} \approx {\rm const_2} \left(\frac{\Dmon}{|V|}\right)^{\frac{\StabIndex-1}{\StabIndex+1}}
\end{equation}
with
\begin{equation}
{\rm const_2}\approx
\Big({3\pi^2}\Big)^{\frac{1-\alpha}{1+\alpha}}.
\end{equation}
For $\StabIndex = 2$, this scaling agree very well with the numerical calculations \cite{Schreiber1982,koehler1989}, as well as with the analytical result based on the coherent potential approximation \cite{Boukahil1990}.
We will denote the scaling laws in Eqs.~\eqref{eq:scaling_N*} and \eqref{eq:width_scaling} as {\it conventional} scaling laws.

For increasingly large values of the disorder, correspondingly strong localization effects will occur. Eventually, the localization length will tend to unity and the exciton states will be essentially localized on one molecule each. Since then the exciton states resemble the monomer excited states to a large extent, it can be trivially deduced that in this limit the exciton absorption spectrum will tend to the monomer absorption spectrum. In particular, we have $\Dagg/\Dmon\approx 1$.

\subsection{Segmentation caused by outliers}
\label{sec:outlier}
We have already discussed the importance of outliers in the case of heavy tailed distributions, i.e.\ small $\StabIndex$, in section \ref{sec:weak_disorder}. 
In this section we will investigate the occurrence of outliers and their interplay with the conventional scaling law in more detail.

As an outlier, we define a site with an energy whose absolute magnitude is larger than $b$ times the coupling $|V|$ between neighboring sites.

The probability for a site to be an outlier is  \mbox{$
\Pout=1-\int_{-b|V|}^{b|V|} \p_{\StabIndex,\ScalePar}(E)\, {\rm d}E
$}.
For $\sigma \ll |V|$ this is approximately given by 
\begin{equation}
\label{eq:def_Pout}
\Pout\approx\frac{2}{\pi}\,\Gamma(\alpha)\, \sin\left(\frac{\pi\alpha}{2}\right)
\left(\frac{\sigma}{bV}\right)^{\alpha}
\end{equation} 
where $\Gamma(\alpha)$ is the 
Gamma function \cite{Abramowitz1964}.
The mean number of outliers in a chain of $N$ sites is thus 
$\overbar{N}_b = N \Pout$. 
This allows us to estimate the mean segment length as
\begin{equation}
\label{eq:mean_N_seg}
\Nseg\approx\frac{N}{\bar{N}_b}=\frac{\pi}{2\,\Gamma(\alpha)\, \sin\left(\frac{\pi\alpha}{2}\right)}
\left(\frac{\sigma}{b|V|}\right)^{-\alpha}
\end{equation} 

With $\Nseg$ and $N^*$ (see Eq.~(\ref{eq:scaling_N*})) we have two competing length scales.
If $N^* \ll \bar{N}_{\rm seg}$ one expects the absorption spectrum to be dominated by states that are localized due to mixing on average on $N^*$ monomers, following the scaling law \eqref{eq:width_scaling}.
On the other hand if $\bar{N}_{\rm seg} \ll N^*$ the spectrum will be dominated by states that are localized on average on $\bar{N}_{\rm seg}$ monomers, but in contrast to the localization due to mixing they correspond to the eigenfunctions of weakly perturbed chains (see \eqref{eq:coeffNN}) with a reduced length of $N \sim \Nseg$.
By equating \eqref{eq:mean_N_seg} and \eqref{eq:scaling_N*} one can estimate when $N^{*} \sim \Nseg$. 

In the following we will choose $b=2$, since from inspection of perfect chains with only one outlier one sees that this already leads to clear segmentation.

\begin{figure}[ptb]
\includegraphics[width=0.95\linewidth]{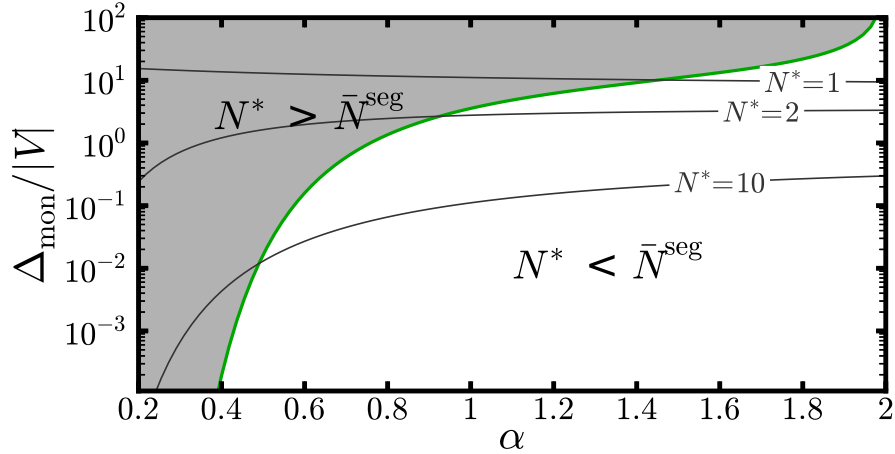}
\caption{\label{fig:segment_phase}
``Phase diagram'' showing the regions in the $(\StabIndex,\Dmon)$-parameter space where different localization mechanisms dominate: the grey area ($\bar{N}_{\rm seg}<N^{*}$) is segmentation- dominated, the white area ($\bar{N}_{\rm seg}>N^{*}$) is dominated by conventional localization. The green line shows $\bar{N}_{\rm seg}=N^{*}$. The gray iso-lines correspond to the $(\StabIndex,\Dmon)$ combinations for which the localization length $N^{*}$ has a certain constant value, given in the plot.}
\end{figure}

Fig.~\ref{fig:segment_phase} shows a ``phase diagram'' that visualizes for which $\StabIndex$ and $\Dmon/|V|$ values one expects either the typical localization length $N^*$ or the mean segmentation length $\Nseg$ to be smaller.
The green line separating the gray and the white region show the combination of $\StabIndex$ and $\Dmon$ for which $N^{*} = \Nseg$ is fulfilled. The gray lines are isolines of  ($\StabIndex$, $\Dmon$) for which $N^{*}$ has a certain constant value.
Fig.~\ref{fig:segment_phase} clearly shows that for small $\StabIndex$ segmentation becomes more important.
For $\StabIndex\approx 1$ one finds $N^{*} = \Nseg$ at $\Dmon/|V| \approx 3.5$. For this combination of $\StabIndex$ and $\Dmon$, the typical localization length is already $<2$ and we are approaching the trivial limit of fully localized states (i.e.\ one site only).
Therefore we do not expect a strong effect of segmentation for $\StabIndex\gtrsim0.9$.
Since both the segment length and the localization length are stochastic quantities with broad distributions an interplay of the two length scales is expected whenever they are of the same order of magnitude.

The expressions for the conventional localization length $N^*$ (Eq.~\eqref{eq:scaling_N*}) and the mean segmentation length $\Nseg$ (Eq.~\eqref{eq:mean_N_seg}) are derived for infinite chains and therefore do not include finite size effects.
The localization properties where $N^*$ and $\Nseg$ are of the same order of magnitude as the chain length $N$ are studied in Fig.~\ref{fig:weak_valid}.
The red lines in Fig.~\ref{fig:weak_valid} are isolines for which the probability $\Pperf$ to find non-segmented chain takes a constant value. With non-segmented chain we mean a single disorder realization of the Hamiltonian \eqref{Ham_ExNN} without the occurrence of an outlier. 
The probability $\Pperf$ for a chain to be non-segmented is (for a given chain length $N$, index of stability $\StabIndex$ and scale parameter $\ScalePar$) determined by $\Pperf = (1 - \Pout)^{N}$, with $\Pout$ the probability to find an outlier, defined by Eq.~\eqref{eq:def_Pout}. 
In Fig.~\ref{fig:weak_valid} isolines for $\Pperf = 99\%$ (solid, red), $\Pperf = 66\%$ (dashed, red) and $\Pperf = 33\%$(dotted, red) are shown. For $\Dmon/|V|$ values below the $\Pperf = 99\%$ line no significant segmentation is expected, whereas for $\Dmon/|V|$ values above the $\Pperf = 33\%$ line segmentation is strongly present.
Whether segmentation due to outliers or localization due to mixing is the dominant effect strongly depends on the index of stability $\StabIndex$. 
Smaller $\StabIndex$ implies that outliers are more probable and therefore segmentation is expected to be more important for small $\StabIndex$. In the following we will again investigate the localization properties of states for different cases of $\StabIndex$, but now with a focus on the regime where the weak disorder limit should be valid according to Eq.~\eqref{eq:weak_disorder_condition}.

As discussed in Sec.~\ref{sec:weak_disorder} the solid black line in Fig.~\ref{fig:weak_valid} marks for which disorder strength $\Dmon/|V|$ the localization length $N^*$ (Eq.~\eqref{eq:scaling_N*}) is equal to the chain length $N$.  For $\Dmon/|V|$ values below that line, the weak disorder limit is valid and for $\Dmon/|V|$ values above that line we are in the conventional scaling regime (see Sec.~\ref{sec:scaling_intermediate_disorder}).

For $\StabIndex = 2$ (left panel Fig.~\ref{fig:weak_valid}) one sees that for all $N$ one needs $\Dmon/|V| > 1$  to find a significant portion of aggregates to be segmented by outliers. For such a large $\Dmon/|V|$  the eigenstates for $N \geq 10$ are already localized due to mixing (i.e., the black solid line lies below the red solid line). Only for very short chains with $N \leq 7$ one finds that the two localization mechanism are on the same order of magnitude. This confirms the conclusion from Fig.~\ref{fig:segment_phase}, that segmentation does not play a significant role for $\StabIndex = 2$.

Similarly, for $\StabIndex = 1$ (center panel Fig.~\ref{fig:weak_valid}) one finds  that in the weak disorder regime (far below the black line) segmentation will not play an important role.

For $\StabIndex = 0.5$ (right panel Fig.~\ref{fig:weak_valid}), as expected, segmentation is prominent in the weak disorder limit. 
For $\Dmon/|V|$, where the localization due to mixing becomes important (solid black line), there are already about $33\%$ of chains segmented. 
While in the conventional scaling regime both localization effects are present, in the weak disorder regime localisation is caused only by segmentation due to outliers.
 On each segment, the weak disorder regime may apply again, but for a reduced chain length that is given by the segment length.
For $\StabIndex = 0.5$ there exists also a (weak disorder) regime where the absorption spectrum Eq.~\eqref{eq:weak_perturb} for the full system size is valid. The validity of Eq.~\ref{eq:weak_perturb} is then given by the probability $\Pperf$ to find a non-segmented chain (from right panel of Fig.~\ref{fig:weak_valid} one sees that very small values of $\Dmon/|V|$ are needed).

\subsection{Dependence on the three regimes on the stability index and the aggregate length. A Summary of the analytical results.}
\label{sec:summary}

\begin{figure}[ptb]
\includegraphics[width=0.95\linewidth]{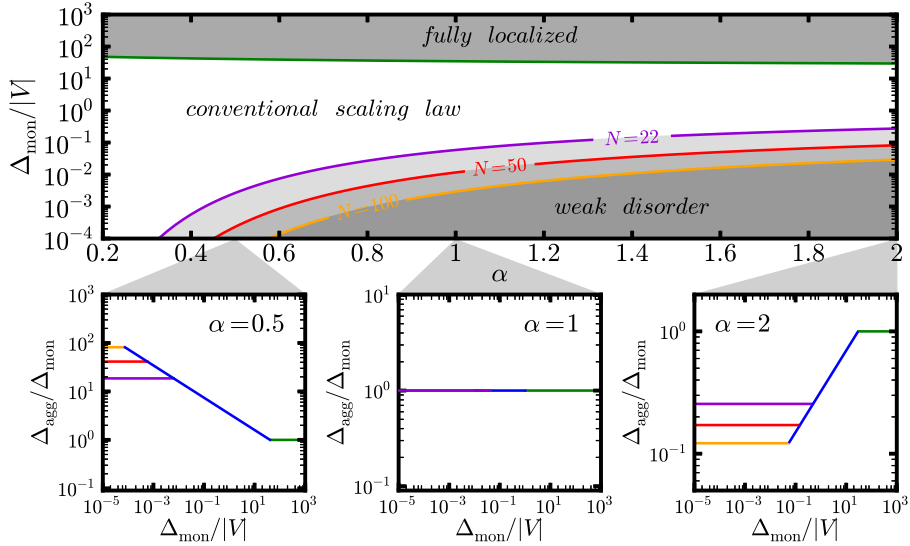}
\caption{\label{fig:fig_regimes_phase_diagram}
(top panel) Phase diagram showing the $(\StabIndex, \Dmon)$ combinations where one expects different types of scalings of $\Dagg$. Colored lines indicate where $N^{*} = 1$ (green) corresponding to the fully localized limit, and $N^*=N$ with chain lengths $N = 22$ (violet), $N = 50$ (red) and $N = 100$ (orange) that demarcate the weak disorder limit. (bottom panels) Expected scaling behavior of $\Dagg/\Dmon$ for the cases of  $\StabIndex = 0.5$(bottom left), $\StabIndex = 1$(bottom center) and $\StabIndex = 2$(bottom right). The color coding matches with the top panel. For $\StabIndex = 1$ all lines collapse on top of each other, since no broadening or narrowing occurs.
}
\end{figure}

In the previous sections, we derived scaling laws for the FWHM $\Dagg$ of the absorption spectrum of a linear chain with nearest neighbor coupling $J$ in three different regimes. Starting from a homogeneous chain (i.e.~no disorder) we used perturbation theory to derive estimates in  the weak disorder limit (see Section \ref{sec:weak_disorder}), where we found a linear scaling of $\Dagg$ with respect to $\Dmon$,
\begin{equation}
\label{eq:D_agg_weak_div_D_mon}
 \frac{\Dagg^{\rm weak}}{\Dmon} = g_{11}(N, \StabIndex) \propto N^{\frac{1}{\StabIndex} - 1}
\end{equation}
with $g_{11}(N, \StabIndex)$ defined by eq.~\eqref{eq:g_approx}. This approximation is valid as long as the basis states do not experience significant disorder-induced mixing. As a rough estimation we found $\Dmon/|V|\ll 3 \pi^2 N^{-\frac{1}{\alpha}}  (N+1)$ (see Eq.~\eqref{eq:weak_disorder_condition}). Increasing the disorder strength will mix low lying states, since they are energetically the closest to each other, resulting in eigenstates that are localized only over $N^*$ monomers. Using the same condition we used for the validity of the weak disorder limit, but for a reduced length $N^*$ of the aggregate, we predict a scaling of the mean localization length $N^*$ (eq.~\eqref{eq:scaling_N*})
\begin{equation}
N^* \propto \left(\frac{\Dmon}{|V|}\right)^{-\frac{\StabIndex}{\StabIndex+1}}.
\end{equation}
Inserting this result in the weak disorder scaling, the relative width of spectra $\Dagg/\Dmon$ does now dependent on the monomer width $\Dmon$ (eq.~\eqref{eq:width_scaling})
\begin{equation}
\label{eq:D_agg_intermediate_div_D_mon}
 \frac{\Dagg^{\rm conventional}}{\Dmon} \propto \left( \frac{\Dmon}{|V|}\right)^{\frac{\StabIndex-1}{\StabIndex+1}}.
\end{equation}
Increasing the disorder strength further will eventually lead to $N^* \approx 1$.
Since the wavefunctions cannot be localized on a length scale smaller than $1$,  the scaling law will not be valid anymore and a spectrum equivalent to that of $N$ uncoupled monomers is obtained,
\begin{equation}
\label{eq:D_agg_strong_div_D_mon}
 \frac{\Dagg^{\rm strong}}{\Dmon} = 1.
\end{equation}
These boundaries between the three regimes allows for the construction of a phase diagram that shows in what part of the parameter space each regime applies. Note that the scaling behavior within the regimes still depends on $\alpha$. This phase diagram is shown in the top panel of Figure \ref{fig:fig_regimes_phase_diagram}, for various values of the system size.
The borders of these regimes are obtained by equating $N^* = N$ for the transition from weak disorder to the intermediate disorder regime (where the scaling laws apply), with chain length $N=22$ (violet), $N = 50$ (red) and $N=100$ (orange), and by equating $N^* = 1$ for the transition of the conventional scaling law to the fully localized regime (strong disorder), marked with the green line.
The bottom panels of Fig.~\ref{fig:fig_regimes_phase_diagram} show examples of the expected scaling behavior for the relative width of the absorption spectrum $\Dagg/\Dmon$ from equations \eqref{eq:D_agg_weak_div_D_mon}, \eqref{eq:D_agg_intermediate_div_D_mon} and \eqref{eq:D_agg_strong_div_D_mon}.
Three representative cases, exhibiting different scaling behavior, are chosen: $\StabIndex = 0.5$ for broadening (bottom left panel), $\StabIndex = 1$ for constant width (bottom center panel) and $\StabIndex = 2$ for narrowing (bottom right panel).

It should be noted that the above considerations on the scaling laws do not include the effect of outliers nor the absorption into energetically higher lying states in the exciton band (Eq.~\eqref{eq:eigVal} with  $j>1$).

\section{Numerical results}
\label{sec:numerics}

In this section we compare the analytical results of section \ref{sec:analytical_estimates} with numerical simulations.
In Sec.~\ref{sec:num_localization_length} we will first study the typical localization length $N^*$ in the various regimes.
In particular we verify numerically the scaling of $N^*$ in the conventional scaling regime. In this context, we will also look at localization length distributions $P(N^{\rm loc})$, since this quantity shows additional signatures of the segmentation effect due to outliers.

After we have analyzed the localization properties of the states that are relevant for absorption, we will turn our attention to the absorption spectra in Sec.~\ref{sec:fwhmscaling}.
First, we will present the numerically obtained scaling of the FWHM $\Dagg$ with respect to $\Dmon$ of the absorption spectrum for different chain length $N$ and index of stability $\StabIndex$. We compare the numerical results with the analytically predicted scaling behavior for all three regimes: weak disorder, conventional scaling and the fully localized regime.
We generally find good agreement between the analytical theory and numerical simulation, however there are some interesting additional features not present in the analytical prediction. These additional features are related to the occurrence of finite size effects and segmentation of the aggregate due to outliers.

To analyze these additional features, we look in detail at the wave functions, the magnitude of the  disorder-induced mixing, and the absorption spectra in Sec.~\ref{sec:scal_50}. We can confirm that the wave functions indeed localize in a way that is consistent with our analytical modeling, in that they form a so-called hidden structure.~\cite{Malyshev1995,Malyshev2001,Malyshev2007,DoMa04_226_,Vlaming2009a,Augulis2010} In addition, there can be contributions to the absorption from higher energy states, modifying the peak widths and leading to additional features mentioned earlier. The occurrence of outliers and the resultant segmentation also significantly contribute to the absorption spectra for small $\StabIndex$.

In the last part of this section, Sec.~\ref{sec:truncated}, we will analyze the effect of truncating the tails of the stable distributions, i.e.\ discarding all values where $|D_n| >b|V|$, with some given $b$. We deliberately choose the same symbol $b$, that we used to define outliers (see Sec.~\ref{sec:outlier}), as the purpose of truncation is to exclude outliers from the distribution.

\subsection{The localization length}
\label{sec:num_localization_length}

\begin{figure*}
 \includegraphics[width=.45\linewidth]{{{deloc_scaling_alpha_2}}}
 \includegraphics[width = .45\linewidth]{{{FWHM_scaling_alpha_2.0}}}
 \caption{\label{fig:scal_2} Scaling of mean delocalization length $\Nloc$ (left panel) and relative FWHM $\Dagg/\Dmon$ (right panel) for $\StabIndex = 2$ for chains length $N = 22$ (orange), $N = 50$ (red) and $N = 100$ (blue). The horizontal gray lines and the diagonal black line mark the predicted scaling for the three different regime (see Eqs.\eqref{eq:scaling_N*}, \eqref{eq:D_agg_weak_div_D_mon}, \eqref{eq:D_agg_intermediate_div_D_mon} and \eqref{eq:D_agg_strong_div_D_mon}). For the scaling of $\Dagg/\Dmon$ the prefactor of the conventional scaling ${\mathrm const_2}$ in Eq.~\eqref{eq:width_scaling} has been adjusted to ${\mathrm const_2} = 0.4$ to match the numerical data.}
\end{figure*}

\begin{figure*}
 \includegraphics[width=.45\linewidth]{{{deloc_scaling_alpha_1}}}
 \includegraphics[width = .45\linewidth]{{{FWHM_scaling_alpha_1.0}}}
 \caption{\label{fig:scal_1} As Fig.~\ref{fig:scal_2}, but now for $\StabIndex = 1$ and with ${\mathrm const_2} = 1$. Note that in contrast to Figures \ref{fig:scal_2} and \ref{fig:scal_05} the y-axis for the scaling of $\Dagg/\Dmon$ is linear, since it doesn't span over many orders of magnitude.}
\end{figure*}

\begin{figure*}
 \includegraphics[width=.45\linewidth]{{{deloc_scaling_alpha_0.5}}}
 \includegraphics[width = .45\linewidth]{{{FWHM_scaling_alpha_0.5}}}
 \caption{\label{fig:scal_05} As Fig.~\ref{fig:scal_2}, but now for $\StabIndex = 0.5$ and with ${\mathrm const_2} = 3$.}
\end{figure*}

From the analytical estimates of the previous section \ref{sec:analytical_estimates}, we have seen that the localization length  of the exciton states (whose typical value we denoted by $N^{*}$) is an important quantity. In this section, we will discuss both the mean localization length  as well as the distribution of localization lengths.
As mentioned in Sec.~\ref{sec:model_hamiltonian}, we will use the inverse participation number to numerically quantify the localization length.

To estimate $N^*$ numerically we use the localization lengths $N^{\rm loc}_j$ (see eq.~\ref{eq:PN}) of states $\ket{\phi_j}$ that are relevant for absorption, that is, those in the vicinity of the lower exciton band edge ($E=-2|V|$ for negative nearest-neighbor interactions $V$).
However, these energies show a spread that depends sensitively on the disorder strength $\Dmon/|V|$. 
One therefore needs a prescription which states one wants to take into account to gain information on $N^*$.

Here, we follow Ref.~\onlinecite{KlMaKn08_084706_,Vlaming2009a} and take all states within a certain energy interval  $[E_{min},E_{max}]$ located in the central region of the absorption spectrum into account. In this approach, we scale the interval with disorder strength in such a way that it is broad enough to cover all relevant s-like states (i.e.~those states without nodes which will thus give a large contribute to the absorption) without containing too large a fraction of mostly dark states.
We  scale the energy interval  according to a transformation of the form
\begin{equation}
\label{eq:interval_loc}
E(\tilde E) = E_0 - a \; \ScalePar^{\frac{2\StabIndex}{\StabIndex + 1} } + \tilde E \; b \ScalePar^{\frac{2\StabIndex}{\StabIndex + 1} }
\end{equation}
with parameters $a$ and $b$ as well as the scaled energy $\tilde E$ chosen in an appropriate way to follow the FWHM scaling of the conventional scaling law.

For the following calculations on the localization length, we will consider states in a certain interval $[E_{min}, E_{max}]$ where the interval is determined by Eq.~\eqref{eq:interval_loc} with the following parameters
\footnote{For a different choice of parameters $a$, $b$, $E_0$, $\tilde E_{min}$ and $\tilde E_{max}$ one obtains a different interval $[E_{min}, E_{max}]$. If this interval covers too many higher lying states ($j\gg1$) one overestimates the mean localization length, since these states are not as much localized as the optical dominant states at the band edge. For $\StabIndex\lesssim0.5$ however a larger interval $[E_{min}, E_{max}]$ allows to see individual segments in the distribution of localization length as small peaks, corresponding to perfectly delocalized states of a segment with length $N^{\rm seg}$. With the parameters used in the present work, we found a balance between a not too small and not too large interval.}
: $a = 0$, $b = 3$.
As the scaled interval $[\tilde E_{min}, \tilde E_{max}]$ we choose $\tilde E_{max} = -\tilde E_{min} = 0.5$. The center of the interval $E_0$ is chosen as the energy of the  $j=1$ state of the corresponding disorder-free chain (see eq.~\eqref{eq:eigE}).

In the left panels of Figures \ref{fig:scal_2} - \ref{fig:scal_05} the mean localization length $\Nloc$, extracted from the numerical simulations, is shown for various disorder strengths $\Dmon$ and chain lengths $N$. 
The individual panels correspond to three different cases of the index of stability $\StabIndex = 2$ (Fig.~\ref{fig:scal_2}), $\StabIndex = 1$ (Fig.~\ref{fig:scal_1}) and $\StabIndex = 0.5$ (Fig.~\ref{fig:scal_05}).
The black lines in the panels correspond to the analytically estimated scaling of the localization length $N^*$ derived in eq.~\eqref{eq:scaling_N*}.
For all cases of $\alpha$, we can clearly observe the three different regimes: weak disorder, conventional scaling and strong localization.
For small disorder values $\Dmon/|V|$, a constant localization length is observed, which simply reflects a full delocalization over the finite size of the chains. The numerical value can be determined analytically by evaluating the inverse participation number Eq.~\eqref{eq:PN} for the solution of the disorder-free aggregate (Eq.~\eqref{eq:coeffNN}), yielding $N^{\rm loc} = \frac{2}{3}(N+1)$ (for $N>1$).
Likewise, the emerging plateau for large disorder strengths $\Dmon/|V|$ reflects  full localization on a monomer.
In between these two extremes, in the intermediate regime, the numerical results match very well with the predicted scaling relation for $N^*$.
The exponent $-\frac{\StabIndex}{\StabIndex+1}$ in eq.~\eqref{eq:scaling_N*} is reproduced perfectly (i.e.~the slope on a double-logarithmic scale).

The prefactor $\mathrm{const}_1$ in eq.~\eqref{eq:scaling_N*} matches the numerics very well for $\StabIndex = 2$ (Fig.~\ref{fig:scal_2}) and it gives the correct order of magnitude for $\StabIndex = 1$ (Fig.~\ref{fig:scal_1}) and $\StabIndex=0.5$ (Fig.~\ref{fig:scal_05}).

The perfect agreement of the theoretical and numerical prefactors (i.e.~the offset of the straight line compared to the numerical curve in the double-logarithmic plot) for $\StabIndex = 2$ in the region of conventional scaling in Fig.~\ref{fig:scal_2} is partially coincidence. 
As mentioned before, there is a certain arbitrariness in how to define the width of the monomer spectrum and the relevant width of the disorder distributions; instead of the FWHM, one could for example also use the scale parameter $\ScalePar$ or the standard deviation, which would lead to a somewhat different prefactor $\mathrm{const}_1$. Using for example $\ScalePar$ as the definition of the width, as has been done in our previous work \cite{Eisfeld2010, Eisfeld2012}, the analytical result for $\StabIndex = 2$ in the conventional scaling regime would be off by a factor of $(4\, \sqrt{\ln 2})^{-2/3}$ from the numerical obtained curve, instead of matching to the extent shown in Fig.~\ref{fig:scal_2}. This offset corresponds exactly to the conversion between $\Dmon$ and $\ScalePar$.

\begin{figure}[ptb!]
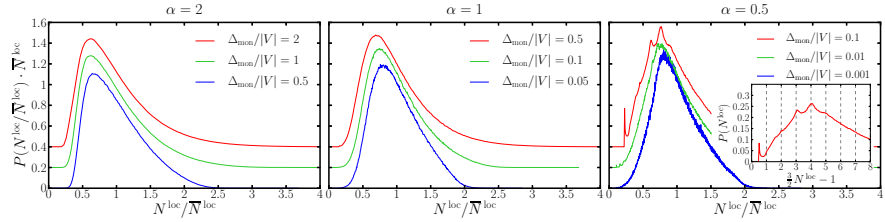

  \includegraphics[width = .9\linewidth]{{{combined_deloc}}}
 \caption{\label{fig:deloc_distrib} Normalized localization length distributions for different $\StabIndex$ and $\Dmon/|V|$. From left to right $\StabIndex = 2$, $\StabIndex = 1$, $\StabIndex = 0.5$. The disorder strength $\Dmon/|V|$ is given in the panels. For better visibility the curves have different offsets. The inset in the right panel shows again the data for $\Dmon/|V| = 0.1$ but with a x-axes scaled in a way that segmentation can be seen more clearly (see text for details).}
\end{figure}

Now, we will take a closer look at the distribution of localization lengths for the three cases of $\StabIndex$ and for different values of $\Dmon/|V|$. 
In particular, we will observe the effect of segmentation due to outliers.
The numerical mean localization length $\Nloc$ does not show a visible effect of segmentation, since it quantifies an entire distribution by a single number.
In this analysis of the localization length distribution, we focus on values of $\Dmon/|V|$ in the intermediate regime, which provides the richest structure. Note that the intermediate regime sets on for smaller values of $\Dmon/|V|$  for increasing aggregate lengths. 
For this regime, a universal shape of the rescaled localization length distribution (i.e., normalized with respect to the average localization length) has been reported for $\StabIndex = 2$ and $\alpha=1$ \cite{Vlaming2009a}.

Figure \ref{fig:deloc_distrib} shows the rescaled distributions of localization lengths $P(N^{\rm loc}/\overbar{N}^{\rm loc})$ for different index of stability.
For $\StabIndex = 2$ and $\StabIndex = 1$ (left and center panel in Fig.~\ref{fig:deloc_distrib}) the distributions exhibit very similar shapes independent of the disorder strength $\Dmon$, confirming the previously reported universal behavior.~\cite{Vlaming2009a}

For $\StabIndex = 0.5$ (right panel in Fig.~\ref{fig:deloc_distrib}) an additional structure on top of an otherwise universal shape can be seen, destroying the universality of the distributions.
This structure starts to visibly appear for $\Dmon/|V|\approx10^{-2}$ and becomes more prominent with increasing disorder strength
\footnote{If one increases the interval $[E_{min}, E_{max}]$ from which eigenfunctions are taken and the localization length $N^{\rm loc}$ is calculated, then visible small peaks will already appear for $\Dmon/|V| < 10^{-2}$}. 
This additional structure is a signature of segmentation, where the part of the chain between two outliers effectively acts as a decoupled chain of reduced length.
Note, that the $\Dmon/|V|$ values used in Fig.~\ref{fig:deloc_distrib} shown are well within the conventional scaling regime (compare left panel of Fig.~\ref{fig:scal_05}).
All the additional peaks in the localization length distribution can be traced back to states with a localization length of $N^{\rm loc} = \frac{2}{3}(N^{\rm seg} + 1)$, with $N^{\rm seg}$ being the length of an aggregate with the respective reduced length. This is best seen in the inset in the left panel of Fig.~\ref{fig:deloc_distrib}, where we have used an inverse of 
the above formula $N^{\rm seg} = \frac{3}{2} N^{\rm loc} - 1$ to scale the distributions $P(N^{\rm loc})$ to $P(N^{\rm seg})$. One can clearly see that all the additional peaks in the localization length distribution appear at integer values of $N^{\rm seg}$, indicating segments of reduced length.

As we have seen, the numerical mean localization length $\bar N^{\rm loc}$ follows the scaling of the analytical estimation of $N^*$ in the conventional scaling regime remarkably well, even in the presence of segmentation. 
Therefore we expect that also the FWHM $\Dagg$ of the main absorption peak still follows the predicted scaling law $\Dagg/\Dmon \propto (\Dmon/|V|)^{\frac{\alpha - 1}{\alpha - 1}}$ to a large extent.
In the next section, the scaling of the absorption peak width is studied in detail.

\subsection{Scaling of the FWHM}
\label{sec:fwhmscaling}

As argued in Sec.~\ref{sec:analytical_estimates}, the width of the absorption spectrum can theoretically be linked to the localization behavior of the exciton states at the bottom of the band. In this section, we will directly test the resulting predictions by numerically calculating absorption spectra, for various chain lengths $N$ and for different values of the index of stability $\StabIndex$ and comparing the peak widths with the analytical expressions derived in Sec.~\ref{sec:analytical_estimates}. The general agreement between the analytical predictions and the numerical results is good, although a few additional features arise on top of the predicted scaling. The origin of these additional features is discussed in Sec.~\ref{sec:kink_jigsaw}. 
Details on the numerical procedure can be found in \ref{app:num_proc}.

The numerical results for the FWHM calculation are shown in the right panels of Figures \ref{fig:scal_2} -- \ref{fig:scal_05}, where we have plotted the scaling $\Dagg/\Dmon$ for $\StabIndex = 2$ (Fig.~\ref{fig:scal_2}), $\StabIndex = 1$ (Fig.~\ref{fig:scal_1}) and $\StabIndex = 0.5$ (Fig.~\ref{fig:scal_05}). The depicted chain lengths are $N=22$ (yellow line), $N=50$ (red line) and $N = 100$ (blue line).
In addition to the numerical data, the predicted scalings Eq.~$\eqref{eq:D_agg_weak_div_D_mon}$, $\eqref{eq:D_agg_intermediate_div_D_mon}$ and $\eqref{eq:D_agg_strong_div_D_mon}$ are shown.
For $\StabIndex = 1$ all the analytical scalings collapse onto one line, making them indistinguishable.

Note, that in Figs.~\ref{fig:scal_2} and \ref{fig:scal_05} we use a double logarithmic representation, while in Refs.~\onlinecite{Eisfeld2010,Eisfeld2012} we have used a linear scale.
The double logarithmic representation allows to see details, in particular for very small values of $\Dmon/|V|$.
Furthermore the conventional scaling law is then just represented by a straight line. 

As expected from the analytical estimates and analogous to the localization length simulations presented in Sec.~\ref{sec:num_localization_length}, also for the FWHM we again clearly observe three regimes: weak disorder where finite size effects dominate, intermediate disorder where we can derive a scaling relation for the FWHM, and strong disorder where the excitons are essentially localized on one molecule each.
The plateau for the very weak disorder case is determined exactly by equation $\eqref{eq:D_agg_weak_div_D_mon}$ (gray lines), and since it essentially reflects fully delocalized excitons it does, of course, depend on the size of the aggregate.
In the intermediate disorder regime, 
the analytical theory accurately predicts the exponent of the scaling law $\frac{\StabIndex - 1}{\StabIndex + 1}$ (i.e.~the slope in the double-logarithmic plot), but does not correctly predict the magnitude of the prefactor (i.e., the vertical offset in the double-logarithmic plot). 
In Figures \ref{fig:scal_2} -- \ref{fig:scal_05}, the factor $\mathrm{const}_2$ in Eq.~\ref{eq:width_scaling} has been adjusted by hand to 
match the numerical data as follows, $\mathrm{const}_2 = 0.4$ for $\StabIndex = 2$, $\mathrm{const}_2 = 1$ for $\StabIndex = 1$ and $\mathrm{const}_2 = 3$ for $\StabIndex = 0.5$.
Finally,  in the strong disorder regime the numerical results agree quite well with the analytically predicted plateau at $\Dagg/\Dmon = 1$.

For $\StabIndex = 2$ (right panel of Fig.~\ref{fig:scal_2}), the well known exchange narrowing effect is observed. For small values of the disorder, the exciton peak width increases proportionally to the disorder strength, reflecting the constant, fully delocalized nature of the excitons. Intermediate values of the disorder show excellent agreement with the predicted scaling exponent. At the transition from the weak coupling regime to this conventional scaling regime there is a small, hardly visible ``bump'' appearing, whose origin is discussed in Sec.~\ref{sec:kink_jigsaw}. Finally, large disorder values eventually lead to completely localized states, where $\Dagg\approx\Dmon$.

For $\StabIndex = 1$ (right panel of Fig.~\ref{fig:scal_1}), the scaling exponent of $\Dagg/\Dmon$ vanishes and we expect no change in the relative width of the spectrum. 
This condition is largely fulfilled, except for an increase of roughly $10\%$ in the aggregate width compared to the monomer width in the region of the conventional scaling.\footnote{Since the scaling of $\Dagg/\Dmon$ for all three regimes collapse on a single line, the regimes can not be identified directly. However, from the scaling of the mean localization length (left panel of Fig.~\ref{fig:scal_1}), the three different regions can be distinguished.}
This ``bump'' depends  on chain length and can thus be attributed to finite size effects. 
In the next section, we show that this ``bump'' has the same origin as the ``bump'' in the scaling for $\StabIndex = 2$, described above.
The deviation from $\Dagg/\Dmon = 1$ in the weak disorder regime is due to the numerical error in determining the width of the absorption spectra as well as noise in the absorption spectra due to finite number of realizations ($10^{7}$ for Figures \ref{fig:scal_2} -- \ref{fig:scal_05}).
These small deviations are also present for $\StabIndex = 2$ and $\StabIndex = 0.5$, however they cannot be resolved on the scale presented in Figures \ref{fig:scal_2} and \ref{fig:scal_05}.

For $\StabIndex  = 0.5$ (right panel of Fig.~\ref{fig:scal_05}) the analytically expected broadening of the relative width is observed.
Here too a ``bump'' appears at the transition from weak disorder to the conventional scaling regime.
The ``bump'' is again a finite size effect, though its origin differs from $\StabIndex = 2$ and $\StabIndex = 1$.
In addition, there is a faint \jigsaw{} structure for $\Dmon > 10^{-1}$, which exhibits a smooth transition to the fully localized regime with $\Dagg/\Dmon = 1$. This structure seems independent on the size of the aggregate and  becomes more pronounced when looking at even smaller $\StabIndex$ (see top panel of Fig.~\ref{fig:alpha03}). The ``bump'' as well as this \jigsaw{} structure can be traced back to a frequent occurrence of outliers, as will be discussed in Sec.~\ref{sec:scal_50}.

\subsection{Disordered wave functions and additional structure in absorption spectra}
\label{sec:scal_50}

In the previous section we demonstrated that the numerically obtained scaling of the relative width $\Dagg/\Dmon$ mostly follows the analytically predicted scaling. However, additional features arise, such as a bump in the absorption peak width in the transitional region where localization becomes important, and a fine structure for certain parameters. These features and the general nature of the underlying wave functions that lead to the observed behavior are explained in this section.

We will discuss the wave functions and absorption spectra in the weak disorder regime, the conventional scaling regime and the fully localized regime separately, comparing the three exemplary cases of $\StabIndex = 2$, $1$ and $0.5$ within these regimes. 
To underline our statements we will present absorption spectra for different $\Dmon/|V|$.

\begin{figure*}
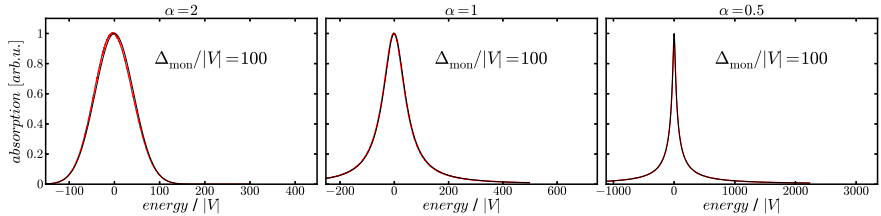

\includegraphics[width = .9\linewidth]{{{combined_strong_absorption_spectra}}}
\caption{\label{fig:spec_wav_strong} Absorption spectra for $\StabIndex = 2$, $\StabIndex = 1$ and $\StabIndex  = 0.5$ in the fully localized regime. Each panel compares the numerical obtained absorption spectra (red) with the analytical absorption spectra (black line) given by the bare uncoupled (i.e. $V = 0$) chain of monomers $\mathcal{A}(E) = N\;\p_{\StabIndex,\ScalePar}(D_n)$.}
\end{figure*}

In Figures \ref{fig:spec_wav_weak}, \ref{fig:spec_wav_con_disorder_2}--\ref{fig:spec_wav_con_disorder_05} and \ref{fig:spec_wav_strong}, we show absorption spectra for different $\Dmon/|V|$. 
The disorder strength $\Dmon/|V|$ is chosen to either optimally show certain features, or just to represent a typical absorption spectrum in the corresponding regime. 
In some of the absorption spectra small arrows mark eigenenergies \eqref{eq:eigE} for specific combinations of chain length $N$ and eigenstate $j$.

For each absorption spectrum in Figures \ref{fig:spec_wav_con_disorder_2}--\ref{fig:spec_wav_con_disorder_05}, eigenfunctions of the Hamiltonian \eqref{Ham_ExNN} for a typical single realization of the disorder is shown.
We only show wave functions in the energy interval most relevant for absorption. 
The coloring of the individual eigenstates indicates a certain absorption strength $A_j$ defined by equation \eqref{eq:A_j}.
A vertical blue line is drawn in the panel of wavefunction, indicating the position of the outlier.

The presented absorption spectra in Fig.~\ref{fig:spec_wav_weak} -- \ref{fig:spec_wav_strong} are used to demonstrate the validity of our analytical assumptions as well as to explain additional features of the numerical scaling of the FWHM from the analytical predicted scaling.
We will show that the deviations originate from finite size effects as well as segmentation due to outliers. Both of these effects are not explicitly considered in the analytical derivations of Eq.~\eqref{eq:D_agg_weak_div_D_mon} and Eq.~\eqref{eq:D_agg_intermediate_div_D_mon}. 
In Fig.~\ref{fig:scal_05} one sees, that the \jigsaw{} structure is roughly independent of the chain length $N$. The bump at the transition from the weak disorder regime to the conventional scaling is $N$-dependent (it occurs at smaller $\Dmon/|V|$ for longer chains), but looks similar for the cases of $N$ shown in Fig.~\ref{fig:scal_05}.

In the following we will exemplarily show results for the case of $N=50$. In Sec.~\ref{sec:kink_jigsaw}, where we will discuss the origin of the ``bump'' for the three different cases of $\StabIndex$, the dependence of the position $\Dmon/|V|$ of the ``bump'' with respect to the chain length $N$ will also become apparent.

\subsubsection{Wave functions: hidden structure and mixing:}
\label{sec:mixing}

\begin{figure*}
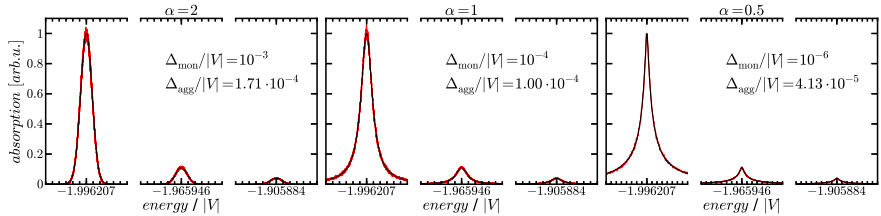

\includegraphics[width = .9\linewidth]{{{combined_weak_absorption_spectra}}}
 
\caption{\label{fig:spec_wav_weak} Absorption spectra  for $\StabIndex = 2$, $\StabIndex = 1$ and $\StabIndex  = 0.5$ in the weak disorder regime. Each panel compares the numerically obtained absorption spectra (red) with the analytical absorption spectra (black line) given by Eq.~\eqref{eq:weak_perturb} for $N=50$. The two curves are nearly indistinguishable. The distance between the ticks on the x-axis is  $\Dagg/2$, i.e.\ half of the FWHM  of the main absorption peak.}
\end{figure*}

In the weak disorder regime, the interpretation of the results is straightforward. The exciton wave functions
are strongly delocalized, and look identical to those of the unperturbed chain (see Eq.~\eqref{eq:coeffNN}); therefore, the homogeneous exciton states are an excellent ansatz for calculating the peak energy and the corresponding oscillator strengths. The peak width is essentially the spread in energy $\Delta_{11}$ of the superradiant peak, and is given by exchange narrowing or broadening over a fully delocalized state. For this regime, in Figure \ref{fig:spec_wav_weak} the numerically obtained absorption spectra are shown together with the analytical predicted absorption spectrum. The latter is simply obtained by calculating the exciton energies and oscillator strengths from the homogeneous expressions Eqs.~\eqref{eq:eigE} and \eqref{F_nu}, and convoluting the peaks with the corresponding $\StabIndex$-distribution of width $\Dagg = g_{11}(N=50, \StabIndex) \Dmon$ 
(see Eq.~\eqref{eq:g_jj'}). As can be seen in Fig.~\ref{fig:spec_wav_weak}, the numerical and analytical spectra match perfectly for all values of $\alpha$.

\begin{figure*}
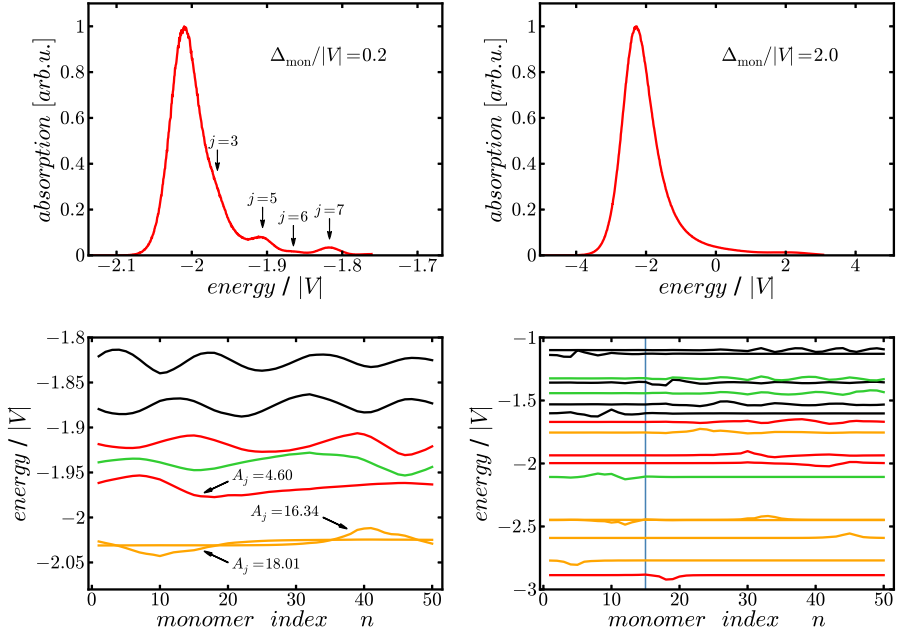

 \includegraphics[width = .45\linewidth]{{{conv_disorder_alpha_2.0_Delta_0.2}}}
 \includegraphics[width = .45\linewidth]{{{conv_disorder_alpha_2.0_Delta_2.0}}}\\
 \includegraphics[width = .45\linewidth]{{{wavefunctions_alpha_2.0_Delta_0.2}}} 
 \includegraphics[width = .45\linewidth]{{{wavefunctions_alpha_2.0_Delta_2.0}}}
\caption{\label{fig:spec_wav_con_disorder_2} Absorption spectra and wavefunction of a single disorder realization with $\StabIndex = 2$ for two different values of $\Dmon/|V|$ [(left) at the transition from the weak disorder regime to the conventional scaling regime, (right) within the conventional scaling regime]. Absorption spectra: Numerically obtained spectra are shown. Arrows mark certain energies, belonging to higher lying eigenenergies $j > 1$ or $j=1$ states for shorter (segmented) chains with $N^{\rm seg} < 50$ of the disorder-free Hamiltonian. Wavefunctions: A typical realization of wavefunction is shown. The coloring of the wavefunctions indicate a certain absorption strength $A_j$ relative to the maximum absorption strength of this particular realization of disorder. $1 \geq A_j/\max(A_j) > 0.5$ (orange), $0.5 \geq A_j/\max(A_j) > 0.2$ (red), $0.2 \geq A_j/\max(A_j) > 0.1$ (green) and $0.1 > A_j/\max(A_j)$ (black).  The vertical blue lines mark the position of outliers defined by $|D_n| > 
2V$ (see Sec.~\ref{sec:outlier}). In the bottom left panel the absolute values of the absorption strengths $A_j$ of particular states are shown.}
\end{figure*}

\begin{figure*}

 \includegraphics[width = .45\linewidth]{{{conv_disorder_alpha_1.0_Delta_0.02}}}
 \includegraphics[width = .45\linewidth]{{{conv_disorder_alpha_1.0_Delta_0.4}}}\\
 \includegraphics[width = .45\linewidth]{{{wavefunctions_alpha_1.0_Delta_0.02}}}
 \includegraphics[width = .45\linewidth]{{{wavefunctions_alpha_1.0_Delta_0.4}}}

\caption{\label{fig:spec_wav_con_disorder_1} Same as Fig.~\ref{fig:spec_wav_con_disorder_2} but now for $\StabIndex = 1$.
}
\end{figure*}

\begin{figure*}
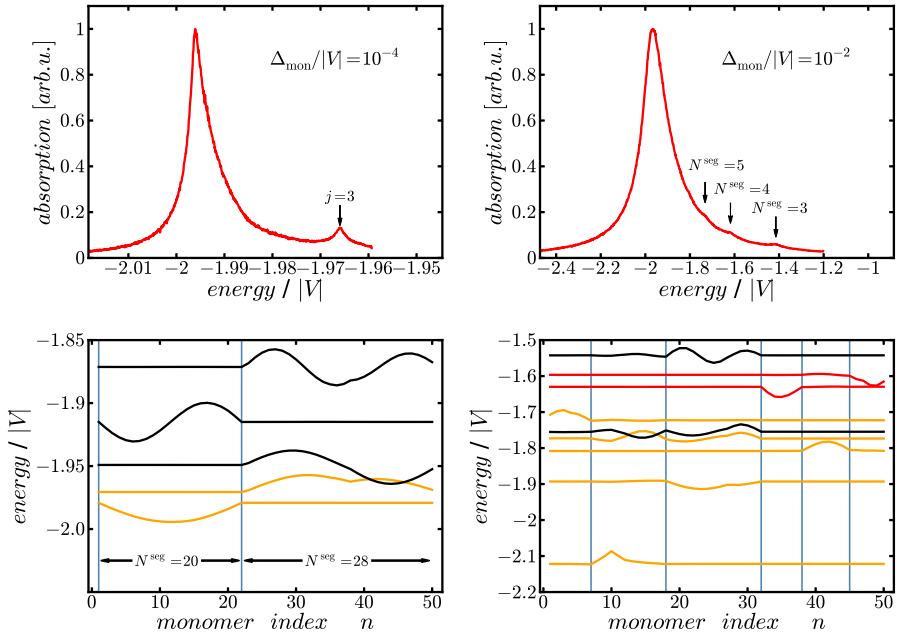

 \includegraphics[width = .45\linewidth]{{{conv_disorder_alpha_0.5_Delta_0.0001}}}
 \includegraphics[width = .45\linewidth]{{{conv_disorder_alpha_0.5_Delta_0.01}}}\\
 \includegraphics[width = .45\linewidth]{{{wavefunctions_alpha_0.5_Delta_0.0001}}}
 \includegraphics[width = .45\linewidth]{{{wavefunctions_alpha_0.5_Delta_0.01}}}
\caption{\label{fig:spec_wav_con_disorder_05} Same as Fig.~\ref{fig:spec_wav_con_disorder_2} but now for $\StabIndex = 0.5$. Note that in the upper right panel the arrows mark peaks belonging to different segment length.
}
\end{figure*}

For increasing amounts of disorder, the lowest exciton states start to mix and as a result, localized states will arise at the bottom of the band. 
At the onset of the intermediate regime, where the localization length of the lowest energy states starts to drop below the chain length, the absorption spectra as shown in the upper left panel of Figures \ref{fig:spec_wav_con_disorder_2}--\ref{fig:spec_wav_con_disorder_05} can still be understood well in terms of the homogeneous exciton solutions Eq.~\eqref{eq:eigVal} and \eqref{eq:eigE}. 
That is, the exciton states still closely resemble the homogeneous solutions (see bottom left panels of Figures \ref{fig:spec_wav_con_disorder_2}--\ref{fig:spec_wav_con_disorder_05}), where the main peak corresponds to absorption into the superradiant $j=1$ exciton state (Eq.~\eqref{eq:coeffNN} with $j=1$), and the smaller peaks at higher energies can be identified with absorption into the exciton states with odd $j>1$. Note that here, while the exact solutions are not quite the disorder-free solutions, we will persist in 
using the labeling from lowest to higher energies by increasing $j$, also to emphasize the resemblance of the disordered wave functions to their disorder-free counterparts.

A larger disorder strength will lead to increased amounts of mixing, i.e.~stronger deviations of the wave functions from the homogeneous solutions and correspondingly stronger localization. As a result, also the states with an even $j$ may gain some oscillator strength. The nature of the mixing is somewhat different for differing values of $\alpha$. For $\alpha=2$, the mixing leads to some oscillator strength in the even $j$-peaks, which although weak may still be resolved in the numerical absorption spectra.
For smaller values such as $\alpha=1$, the mixing as given by the coupling element  $W_{jj'} \propto \Dmon^{2\StabIndex/(\StabIndex + 1)}$ (introduced in section \ref{sec:analytical_estimates}) is larger and such even states tend to strongly mix with the lower $j$ states, giving localized states that do no longer resemble the original even-$j$ states.
 As indicated in the bottom right panels of Figures \ref{fig:spec_wav_con_disorder_2}-\ref{fig:spec_wav_con_disorder_05}, we observe 
increasing amounts of localization near the lower exciton band edge with increasing disorder.
Note that the localized states still have a structure that is reminiscent of the disorder-free solutions, although occurring on a smaller length scale. That is, the lowest energy exciton state in such a localization region is an $s$-like state (i.e.\ it has no nodes), the second lowest state is $p$-like (one node), et cetera. This is referred to as the hidden structure~\cite{Malyshev1995,Malyshev2001,Malyshev2007,DoMa04_226_,Vlaming2009a,Augulis2010}, and the observation of this type of localization behavior is consistent with the estimates made in Sec.~\ref{sec:analytical_estimates} for the scalings of the localization lengths and FWHM with the disorder strength.

For $\StabIndex = 0.5$ we also observe absorption to the higher order odd-$j$ states (see upper left Fig.~\ref{fig:spec_wav_con_disorder_05}). Furthermore we see an asymmetrical broadening of the main absorption peak to the high energy side. This broadening can be attributed to a significant occurrence of outliers.
Distributions with small $\StabIndex$ are characterized by heavy tails and a correspondingly high probability of outliers. 
Such an outlier supports a strongly localized state, but in addition effectively separates the chain into two almost decoupled subchains. On each of these subchains, states resembling the homogeneous solutions (but on a reduced length $N^{\rm seg}$) can occur. 
For small disorder values, this resemblance is strong, and absorption will occur into exciton states that strongly resemble the superradiant $j=1$-state, localized on subchains with a shorter effective length $N^{\rm seg}$. Such states will occur at higher energies for shorter segments, effectively leading to a high energy tail onto the main absorption peak, consisting of $s$-like states on subchains with a distribution of lengths $N^{\rm seg}<N$.
For larger disorder values, increased amounts of mixing occur and a hidden structure will appear on these subchains as well. In the following section, we will illustrate this effect by analyzing disorder distributions with $\alpha=0.3$, where very strong segmentation effects occur and the above features are resolved very clearly.

Finally, very strong disorder leads to exciton states which are very localized, and essentially centered around one molecule. The absorption spectrum is trivial in this case, as the close resemblance of the exciton wave functions to the molecular wave functions leads to an exciton absorption peak which is essentially identical to the monomer absorption peak or equivalently the disorder distribution itself.

As mentioned before, there are additional features present in the dependence of the FWHM on the disorder strength. These will be explained in the next section.

\subsubsection{Additional structure in FWHM: bump and \jigsaw{} structure:}\label{sec:kink_jigsaw}

\begin{figure}
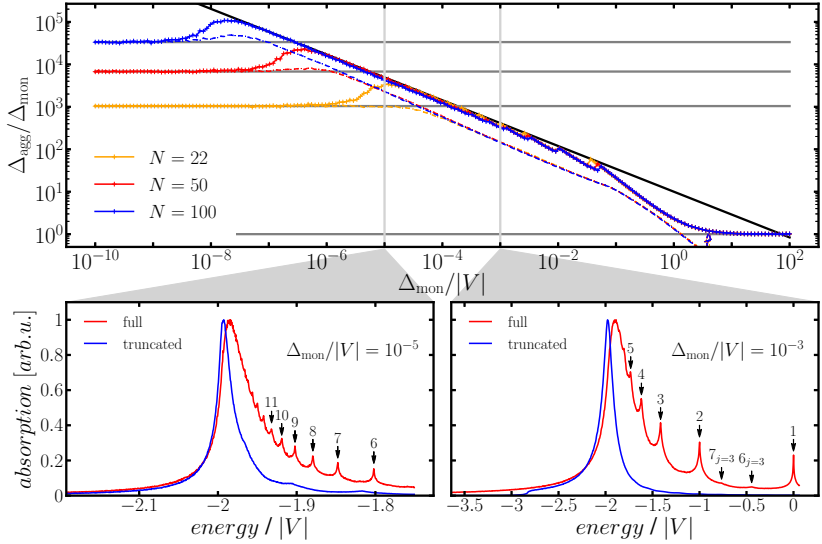

 \begin{center} 
 \includegraphics[width = .9\linewidth]{{{conv_disorder_alpha_0.3_combine}}} 
 \end{center}
 \caption{\label{fig:alpha03} (top panel) Scaling of the relative FWHM $\Dagg/\Dmon$ for $\StabIndex = 0.3$ for 3 different chain lengths. The dashed line corresponds to a truncated distribution with $b=2$ (see Sec.~\ref{sec:truncated}). (bottom panels) Absorption spectra for $\StabIndex = 0.3$ and $N=50$ for a L\'evy-stable disorder (red line) and truncated L\'evy-stable disorder (blue line). The inset shows the position of the spectra in the scaling of $\Dagg/\Dmon$. The arrows at the red line point to energies corresponding to the $j=1$-state for different segment length $\Nseg$ (For $\Dmon/|V| = 0.001$ also $j=3$-states are shown. These are indicated by a subscript.). The arrows at the blue line mark higher excited states $j>1$ for $\Nseg = N = 50$.}
 
\end{figure}

As mentioned before, in the right panels of Fig.~\ref{fig:scal_2} - \ref{fig:scal_05} are two additional features present, namely i) a bump in the transition between the delocalized regime and the intermediate disorder regime, and ii) a \jigsaw{} structure for small values of $\alpha$ (weakly visible for $\alpha=0.5$). Both  features can be understood by examining the wave functions and the mixing behavior, analogous to the analysis presented in Sec.~\ref{sec:mixing}.  

We first analyze the bump that occurs in the peak width in the transitional regime. Besides the main peak that results from absorption into the superradiant $s$-like state, i.e. approximately the homogeneous solution Eq.~\eqref{eq:coeffNN} for $j=1$, there are smaller peaks corresponding to the higher order odd states.
A close look at the peak positions and peak heights at the disorder values where the bump occurs, confirms that for $\alpha=2$ and $\alpha=1$ this deviation from our theoretical estimates is caused by a shoulder in the absorption peak, caused by the $j=3$ transition. 
The additional width provided by absorption into this state, which has not been accounted for in the absorption peak width estimate in Sec.~\ref{sec:scaling_intermediate_disorder}, causes this feature.
This explanation can be confirmed by the fact that it allows for an estimate for the disorder values for which we expect the bump to occur. 
This is for $\Dmon/|V|$ values where the energy gap between the $j=3$ and $j=1$ state, which can be calculated using Eq.~\eqref{eq:eigE}, is equal to the analytically obtainable peak width $\Delta_{11}$ (Eq.~\eqref{perturb_breite}).
This yields estimates of the bump position of $\Dmon\approx0.2|V|$ for $\alpha=2$ and $\Dmon\approx0.03|V|$ for $\alpha=1$, in good agreement with the numerically obtained bump positions of respectively $\Dmon\approx0.27|V|$ and $\Dmon\approx0.04|V|$.
The energy gap between the $j=1$ and $j=3$ state depends on the chain length $N$ and (using Eq.~\eqref{eq:eigE}) is approximately depending on $N$ by $E_{3} - E_{1} \propto N^{-2}$. As the chain length increases the energy gap becomes smaller and a smaller $\Dmon/|V|$ is sufficient for the $j=1$ and $j=3$ peak to overlap. Therefore, for $\StabIndex = 2$ and $\StabIndex = 1$ the bump in the transitional regime will occur at smaller $\Dmon/|V|$ as the chain length increases.

For $\alpha=0.5$, the origin of the bump is similar but not quite identical. Again, the bump is observed to originate from additional peaks growing into the main absorption peak. However, for smaller $\alpha$-values, such as $\alpha=0.5$, these additional peaks correspond to absorption into $s$-like states located on short (outlier-induced) segments instead. Since the absorption for shorter segments occurs at higher energies, this leads to a shoulder on the high energy side and a corresponding asymmetry in the absorption lineshape.
For increasing chain length $N$ the bump appears for smaller values $\Dmon/|V|$. This can be explained by looking at the probability of finding segmented chains. The probability $P_{\geq1}$ to have at least one outlier in a chain of length $N$ is determined by $P_{\geq1} = 1 - (1-\Pout)^{N}$ with $\Pout$ being the probability for an outlier on a single site, defined by Eq.~\eqref{eq:def_Pout}. For small values of $\Dmon/|V|$ we can approximate  $P_{\geq1}$ by $P_{\geq1} \approx N\Pout \propto N\; (\Dmon/|V|)^{\alpha}$. Solving $P_{\geq1}$ for $\Dmon/|V|$ leads to the relation $\Dmon/|V| \propto (P_{\geq1} / N)^{\frac{1}{\alpha}}$. This last relation can be understood as follows. The bump is caused by side peaks on the high-energy side of the main peak due to a significant amount of segmented chain (numerically we find roughly $40\%-60\%$ of segmented chains at the maximum of the bump). 
For large chain length $N$ a smaller value of $\Dmon/|V|$ will lead to the same amount of segmented chains, as a larger 
value $\Dmon/|V|$ does for shorter chain length. This explains the $N$-dependency of the bump.

For small $\alpha$-values and larger disorder strengths, another additional feature is observed: a \jigsaw{} feature on top of the expected scaling behavior.
The origin is, in fact, similar to the origin of the bump explained in the previous paragraphs. With increasing disorder strength, the number of outliers increases and thus an increasing amount of segmentation occurs. This leads to short segments, which in turn support $s$-like states that can be at significantly higher energies than the main peak.
It is easily established at what energies absorption takes place for shorter homogeneous segments; for example, for a segment of length $N=3$ we expect a peak at energy $E\approx-1.41|V|$.
For sufficiently heavy-tailed disorder distributions, these short segments occur frequently enough to lead to  resolvable absorption peaks at the aforementioned energies. 
The \jigsaw{} structure is then nothing more than absorption peaks from shorter segments growing into the main absorption peak. 

The above \jigsaw{} structure is already visible for $\alpha=0.5$ (see right panel of Fig.~\ref{fig:scal_05}), but is not overly pronounced. A similar analysis for a smaller value of $\alpha$ shows the aforementioned effects even more clearly. 
The upper panel of Fig.~\ref{fig:alpha03} shows the disorder dependence of the peak width, while the lower panels of Fig.~\ref{fig:alpha03} show absorption spectra for two disorder values $\Dmon$, all for $\alpha=0.3$. 
The \jigsaw{} features are very clearly visible in the FWHM scaling curve. In addition, one clearly observes a series of peaks on the high energy side of the main absorption peak. It is easily confirmed that the energies of this series of peaks exactly match up with the absorption peaks of different segment lengths. In addition, the absorption spectrum shown in then lower right panel of Fig.~\ref{fig:alpha03},that is in the \jigsaw{} regime, confirms that the additional absorption peaks are growing into the main peak. Finally, it should be noticed that the upper panel of Fig.~\ref{fig:alpha03} shows that the \jigsaw{} feature repeats over increasingly broader disorder intervals. 
This corresponds to the fact that the energy differences 
between a chain of length $N^{\rm seg}$ and length $N^{\rm seg}-1$ increase strongly when $N^{\rm seg}$ gets smaller. For smaller disorder values, the peak width decreases in this \jigsaw{}  fashion by scooping up energetically close absorption peaks of long segments, but for larger disorder values the peak width needs to grow over increasingly large distances to encounter another absorption peak that can grow into the main peak.

\subsection{Truncated L\'evy distributions} 
\label{sec:truncated}
The L\'evy-stable distributions can possess a high probability for the occurrence of arbitrarily large fluctuations.
However, in a real physical system energy fluctuations cannot be arbitrarily large. 
Therefore, it is an important question to what extent the results derived in the previous sections (e.g.\ scaling laws, sawtooth structure etc.) is a result of the heavy tails, and which depend on the central shape of the stable distributions.

In this section we will briefly discuss L\'evy-stable distribution with truncated tails.

\subsubsection{Properties of truncated L\'evy distributions:}
\label{sec:def_trunc}
We define truncated L\'evy-distributions as in Ref. \onlinecite{Mantegna1994}
\begin{equation}
 \label{eq:trunc_levy}
 \p_{\StabIndex,\ScalePar}^{\rm trunc}(D_n) = \mathcal{N}\left\lbrace \begin{array}{ll}
                                \p_{\StabIndex,\ScalePar}(D_n) & , D_n < b|V|\\
                                0 &, \rm else
                               \end{array} \right.                              
\end{equation}
with $\p_{\StabIndex,\ScalePar}(D_n)$ a L\'evy-stable distribution as defined by Eq.~\eqref{fourier} and $b$ the truncation parameter, the value where the distribution is truncated. The normalization factor is given by $\mathcal{N} = 1/(1 - 2\,\int_{b|V|}^{\infty}\p_{\StabIndex,\ScalePar}(D_n)\,dD_n)$. We now refer to  $\ScalePar$ as the scale parameter of the underlying L\'evy-stable distribution. The FWHM $\Dmon$ does in general not have a simple relation to the scale parameter $\ScalePar$ as before. Note that the FWHM is bounded by $\Dmon\leq 2b|V|$ but the parameter $\ScalePar$ is not.

Within the interval $[-b|V|, b|V|]$ the truncated distributions only differ from the L\'evy-stable distribution by the normalization factor $\mathcal{N}$, which scales like $\mathcal{N} - 1 \propto (\ScalePar/(b|V|))^{\StabIndex}$. 
For $\ScalePar/(b|V|) \ll 1$ the normalization factor is on the order of unity, making the truncated and non-truncated distributions nearly indistinguishable. 
A crucial difference between truncated and non-truncated disorder however remains, as truncated L\'evy distributions $\p_{\StabIndex,\ScalePar}^{\rm trunc}(D_n)$ have a finite variance $\gamma^2 = \int\, D_n^2\, \p_{\StabIndex,\ScalePar}^{\rm trunc}(D_n)\, dD_n$.
Therefore a sum of truncated L\'evy distributed random variables $D_n$ (e.g.~the $W_{jj'}$ in Eq.~\eqref{Delta_nunu}) will converge to a Gaussian distributed random variable, according to the central limit theorem. The rate of convergence to the Gaussian limit for truncated L\'evy distributions is briefly discussed in the next section \ref{sec:clt}.

Since the $W_{jj'}$ in Eq.~\eqref{Delta_nunu} are no longer strictly L\'evy-stable distributed we will address the following questions in Sec.~\ref{sec:scal_trunc} \\
(I) Are the derived scalings for the relative FWHM $\Dagg/\Dmon$ Eqs.~\eqref{eq:D_agg_weak_div_D_mon}, \eqref{eq:D_agg_intermediate_div_D_mon} and \eqref{eq:D_agg_strong_div_D_mon} still valid?  \\
(II) What happens to the features caused by outliers?

In consistency with our definition of an outlier in Sec.~\ref{sec:outlier} we set $b=2$.
As an explicit example, we will discuss the truncated distribution for the case of $\StabIndex = 0.3$.
We have chosen this small value of $\alpha$, since the features of the heavy tails can then be seen quite clearly.

\subsubsection{Central limit theorem (CLT) - rate of convergence:}
\label{sec:clt}
In this section we will give a brief overview of the convergence properties of a sum of truncated L\'evy distributed random numbers.

The CLT states in essence\footnote{There exist many versions of the CLT, e.g.~for independent, non-identically distributed random variables. We will only focus on the special case of independent, identically distributed random numbers with zero mean, as those are the distributions used throughout this work.} that, if $X_1, X_2, \dots X_n$ is a sequence of independent, identically distributed random numbers with finite variance $\gamma_x^2$ and mean $\mu = 0$, then the probability distribution $P_n(\sqrt{n}Y_n)$ of the so-called sample mean $Y_n = \frac{1}{n}\sum_{i=1}^{n} X_i$ converges to a Normal distribution with variance $\gamma^2 = \gamma_x^2$ in the limit $n\rightarrow\infty$. 

The CLT makes a statement about the limit $n\rightarrow\infty$, but usually a small number $n$  suffices to be well converged to Gaussian, e.g.~$n\approx5$ for a uniform distribution. 
For a sum of truncated L\'evy distributed random number it has been shown \cite{Mantegna1994,Koponen1995}  that the convergence to a Gaussian distribution strongly depends on the ratio $\ScalePar/(b|V|)$.
For a ratio of $\ScalePar/(b|V|) \ll 1$ the convergence can be very slow, in the sense that 
for $n$ smaller than a critical value $n_0$ the sum of L\'evy truncated random numbers can still be approximated by a L\'evy-stable distribution, while for $n \gg n_0$ the sum can be approximated by a Gaussian random number.
The transition between the two distributions is approximately \cite{Mantegna1994} at $n_0 = \Gamma_\StabIndex (\ScalePar/b|V|)^{-\StabIndex}$, with $\Gamma_\StabIndex$ a prefactor only depending on the index of stability.
The smaller the ratio $\sigma/(b|V|)$, the larger the $n$-interval for which the sum of truncated L\'evy distributed random numbers can be approximated by a L\'evy-stable random number.
Therefore we expect the the derived scaling laws Eq.~\eqref{eq:D_agg_weak_div_D_mon} and Eq.~\eqref{eq:D_agg_intermediate_div_D_mon} to be valid as long as $n_0 \ll \min(N, N^*)$ with $N$ the chain length and $N^*$ the typical localization length (Eq.~\eqref{eq:scaling_N*}).

\subsubsection{Scaling law and absorption spectra for $\StabIndex = 0.3$:}
\label{sec:scal_trunc}

In this section we will address the two questions raised in Sec.~\ref{sec:def_trunc}. We will first check the validity of the scaling laws for the weak disorder Eq.~\eqref{eq:D_agg_weak_div_D_mon}, the conventional scaling regime Eq.~\eqref{eq:D_agg_intermediate_div_D_mon} and finally the fully localized regime Eq.~\eqref{eq:D_agg_strong_div_D_mon}. As expected from Sec.~\ref{sec:clt}, we find in general very good agreement with the analytically derived scaling law for the weak disorder and conventional scaling regime but deviations for the strong disorder regime. This analysis will answer question (I).
In the second part we will analyze the outlier induced features of the scaling laws, namely the bump at the transition from the weak disorder regime to the conventional scaling regime and the sawtooth-shaped structure on top of the conventional scaling.
In consistency with our explanation of these features in Sec.~\ref{sec:kink_jigsaw} we find that by truncating the distributions, and thus excluding outliers, these features almost completely vanish. This will answers question (II).

\paragraph{(I) - Validity of the scaling laws:}

In the top panel of Fig.~\ref{fig:alpha03} the scaling of $\Dagg/\Dmon$ for the cases of $N = 22,\ 50,\ 100$ is shown. The crossed solid lines are the scalings for L\'evy-stable disorder and the dashed line is the scaling for the truncated L\'evy-disorder.
First we compare the three different regimes that we find for the L\'evy stable disorder (see Sec.~\ref{sec:summary}).
For all chain length $N$ the weak disorder limit is the same for the truncated and non-truncated disorder.
Comparing the truncated disorder distributions with the non-truncated one they are almost indistinguishable as $\frac{\sigma}{b|V|} \ll 10^{-6}$. This means that the chosen random variables are almost the same for the two types of disorder and therefore lead to the same spectrum.

The value of the exponent of the conventional scaling law (Eq.~\eqref{eq:D_agg_intermediate_div_D_mon}) is also still valid for the truncated L\'evy-disorder. The difference between truncated and non-truncated distribution is only a constant pre-factor, i.e.\ a constant offset on a double logarithmic scale.
This pre-factor will be explained below.

For the chosen  truncation parameter $b = 2$, there is no strong disorder limit as for the non-truncated case. The curve of the scaling law stops at $\Dmon/|V| =2 b= 4$, since the FWHM $\Dmon$ for the truncated L\'evy disorder distribution cannot exceed the interval $[-b|V|, b|V|]$. 
One finds in addition that for $0.1 \lesssim \Dmon/|V| \leq 4$ the scaling of the truncated $\Dagg/\Dmon$ (dashed line in top panel of Fig.~\ref{fig:alpha03}) does not have the analytically estimated exponent of the conventional law Eq.~\eqref{eq:D_agg_intermediate_div_D_mon} but follows approximately $\Dagg/\Dmon \propto (\Dmon/|V|)^{-1}$. 
This change of the scaling is mainly due to the validity of the CLT (see Sec.~\ref{sec:clt}).
The value of $\frac{b|V|}{\sigma}$ and the localization of the underlying wavefunctions are such that $W_{jj'}$ (Eq.~\eqref{Delta_nunu}) can no longer be approximated by a L\'evy-stable random number of the same index of stability, but is rather Gaussian distributed.
The observed scaling  $(\Dmon/|V|)^{-1}$  indicates that the width of the resulting absorption spectra is independent of $\Dmon/|V|$. This $(\Dmon/|V|)^{-1}$ scaling is observed for all $\StabIndex$ and the individual scaling laws for the different $\StabIndex$ collapse eventually onto one line (not shown). 
The universality of this scaling with respect to the index of stability can be understood, considering that the truncated distribution $\p_{\StabIndex,\ScalePar}^{\rm trunc}(D_n)$ converges to an uniform distribution within the inverval $[-b|V|, b|V|]$ in the limit of $\ScalePar\rightarrow\infty$ independently of $\StabIndex$.

\paragraph{(II) - Segmentation due to outliers:}

Fig.~\ref{fig:alpha03} not only shows the scaling of $\Dagg/\Dmon$ (upper panel) but also two absorption spectra (lower panels). 
First we will discuss the outlier induced features on the scaling laws, before looking at individual absorption spectra.

Comparing the features originating from outliers, namely the bump at the transition from the weak disorder regime to the conventional scaling regime and the \jigsaw{} structure on top of the conventional scaling, one clearly sees in the upper panel of Fig.~\ref{fig:alpha03}, that they have almost vanished for the truncated L\'evy-disorder.
While the \jigsaw{} structure has completely disappeared for all chain length $N$, the disappearance of the bump shows a $N$ dependence.
For $N=22$ the bump has completely disappeared, for $N=100$ however it is only reduced in height.
Changing the parameter $b$ to smaller values (not shown), the height of this bump is further reduced until vanishing completely.
This $N$ dependence can be explained by taking the occurrence of two consecutive outliers with opposite signs into account.
These two outliers create an energy gap of twice the size, compared to just a single outlier. 
For the non-truncated L\'evy-stable disorder it does not matter if one or two consecutive outliers segment the chain, since the actual size of the energy gap does not matter if sufficiently large \footnote{The larger the energy gap the more perfect is the segment, meaning not only a clear $j=1$ state can be found but also states with $j>1$ are clearly localized on this segment. Those $j>1$ states however do not significantly contribute the absorption spectrum.}.
For the truncated L\'evy-disorder their are in principle no single outliers segmenting the chain, but their can be two consecutive large enough fluctuations with opposite sign creating a large enough gap to actually segment the chain.
The probability of finding at least one pair of these consecutive large fluctuations is larger the longer the chain becomes, but still smaller than finding a single large fluctuation. Therefor the bump at the onset still remains as long as the truncation is at sufficient large values, but is reduced in height.

We have shown that the outlier induced features of the scaling of $\Dagg/\Dmon$ vanish for truncated disorder. 
The disappearance of outliers and the resulting segmentation can also be seen in the absorption spectra, where each segment can be seen as a small, but distinct peak.
Looking at individual absorption spectra explains the offset between the truncated scaling and the non-truncated scaling in the conventional scaling regime (upper panel of Fig.~\ref{fig:alpha03}).
The lower panels in Figures \ref{fig:alpha03} show absorption spectra for non-truncated (red line) and truncated (blue line) disorder. 
One immediately notices that the truncated absorption spectrum is narrower than the non-truncated absorption spectrum and the sharp peaks at high energies are missing. 
These sharp high-energy peaks can be clearly assigned to segments with reduced length $N^{\rm seg} < 50$ as indicated by the arrows. 
The sum of all these peaks form a broad shoulder on the high energy side of the main absorption peak, explaining the offset of the scaling $\Dagg/\Dmon$ of the non-truncated disorder in Fig.~\ref{fig:alpha03}, compared to the truncated disorder.
The truncated absorption spectrum also has peaks at the high-energy side. These peaks can be assigned to states with $j = 3$, $5$ and $7$ for a $N=50$ aggregate (Eq.~\eqref{eq:eigE}), as indicated by the arrows. The appearance of peaks with ($j > 1$) have already been discussed for $\StabIndex  = 2$ and $\StabIndex = 1$ in Sec.~\ref{sec:kink_jigsaw}. From Fig.~\ref{fig:alpha03} one can see that the absorption spectra for truncated disorder show similar behavior as the absorption spectra of $\StabIndex = 2$ (Fig.~\ref{fig:spec_wav_con_disorder_2}) or $\StabIndex = 1$ (Fig.~\ref{fig:spec_wav_con_disorder_1}), where outliers do not play a significant role.

We have demonstrated that by truncating the L\'evy-stable distributions at $b = 2$ (Eq.~\eqref{eq:trunc_levy}) the signatures of outliers almost completely vanish. The sawtooth-shaped structure on top of the conventional scaling has completely disappeared and the bump at the transition from the weak disorder regime to conventional scaling regime is significantly reduced in height. The changes in these two features can be traced back to the strongly reduced likelihood of outliers.
Since outliers do almost not occur, the aggregate is very unlikely to split into segments, where short segments would then give rise to additional peaks in the absorption spectrum.

\section{Conclusions} \label{sec:conclusions}

We have provided an in-depth analysis of the effects of static (diagonal)
disorder on the localization and absorption properties of excitonic
systems, where we have generalized from the commonly considered Gaussian
distributions to the wider class of L\'evy stable distributions.
This generalization, allowing for heavy-tailed distributions,
shows a rich palette of novel effects and regimes, owing to the interplay
of various length scales. We provide a comprehensive study of the
localization behavior and its optical implications throughout the entire
parameter regime.

There are three relevant length scales. First of all there is the length of the aggregate $N$.
The second length scale is related to the occurrence of outliers, i.e. sites with transition energies in the tail of the distribution and which as a result couple only weakly to other molecules. 
Outliers effectively lead to a subdivision of the system into weakly coupled segments. 
The mean segment length $N^{\rm seg}$ is the relevant length scale here. 
The disorder also leads to Anderson localization, and
L\'evy stable disorder distributions likewise imply localized states, in
particular at the optically relevant bottom of the band. The typical
localization length $N^{*}$ at the bottom of the exciton band provides the third length scale. It should be
noted that both the segments and the Anderson localization lengths are
described by broad distributions, leading to a non-trivial interplay of
both these effects over a broad region in parameter space.
The fact that the optical response for $J$-aggregates is dominated by the states at the lower
exciton band edge suggests that we primarily focus on the localization
properties of those states.

Several distinct regimes can be distinguished, depending on the relative
magnitude of the above three length scales. First of all, one can identify
a weak disorder regime (small $\Dmon/|V|$). This implies a long localization
length, so that the finite size of the system is a limiting factor for the
spatial extent of the exciton states. In addition, for smaller values of
$\alpha$, outliers are still likely to occur and thereby creating
segments. In that case, it is the segment length that determines the
coherent spread of the excitons. In this regime, the eigenstates strongly
resemble the homogeneous solutions of a chain of length $N^{\rm seg}$ or
length $N$. The absorption is thus dominated by the superradiant $s$-like
states of the various segments, which occur at $E=-2\left|V\right|$ for
long segments and chains, and at somewhat higher energies for shorter
segments.
The absorption peak width exhibits the well known exchange narrowing by a factor of $\sqrt{N}$ for Gaussian disorder.
In general we find that the absorption peak width is scaled by a $N^{1/\StabIndex - 1}$, which leads to a narrowing for $\StabIndex > 1$ and to a broadening for $\StabIndex < 1$.
These analytical predictions are
confirmed numerically, where both the wave function shapes as well as the
absorption peak width agree extremely well with the theoretical forecasts.

For larger values of the disorder strength, the localization length
drops below the chain length, and we can obtain analytical expressions
for the scaling behavior of both the mean localization length and the 
absorption peak width with disorder. Numerically, both these scaling relations are reproduced quite
well, with an excellent agreement with regards to the exponent but some
arbitrariness remaining in the prefactor. In addition, and most
prominently for smaller values of $\alpha$, additional structures appear
superimposed on the numerically calculated localization length
distributions and on the scaling of the absorption peak width. The
localization length distributions for small $\alpha$ show an additional
series of peaks, which can straightforwardly be identified with excitons
localized on (very) short segments. Likewise, a bump at relatively small
disorder strengths and a zigzag structure for larger disorder strengths
and small stability index $\alpha$ can be observed. We have shown that all
these are related to additional peaks growing into the main absorption
peak, with the zigzag structure being yet another clear indicator of the
occurrence of segments.

At large disorder values, one eventually approaches the regime where all
exciton states are strongly localized, with a localization length tending
to 1. The exciton states will then strongly resemble the molecular excited
states, and the absorption peak width will trivially approach the monomer
absorption peak width - which is confirmed numerically as well.

One manifestation of the shape of L\'evy stable distributions is the
increased amount of weight in the tails for decreasing $\alpha$; however,
since arbitrarily large fluctuations (i.e. arbitrarily long tails) may not
be physical, we have also analyzed the behavior for L\'evy stable
distributions with small $\alpha$ with the tails removed. A truncation of
the tail at the exciton band edges, i.e.\ at twice the nearest neighbor
interaction, shows that the same regimes apply as before, set in at
very similar parameter values, and show the same scaling exponents.
However, segmentation becomes much more rare, and as a consequence the
direct signatures of segmentation that were observed before diminish
considerably or vanish completely. This includes the absence of short
segment peaks in both the localization length distributions and absorption
spectra, and the resultant disappearance of the zigzag structure and a
strong reduction in the bump in the absorption peak width curves. In
addition, while the scaling exponent remains the same, the absorption
peaks are more narrow, due to the absence of shorter segment contribution
that tend to occur on the high energy side of the absorption peak.

\appendix
\section{Numerical procedure}
\label{app:num_proc}
We will briefly describe the numerical implementation of the L\'evy-disordered aggregate and our method to extract the FWHM of noisy data.

\paragraph{Hamiltonian, Histograms:}
Independent L\'evy-stable (pseudo) random numbers $D_n$ are generated using the procedure described in Ref.~\cite{Chambers1976}.
For a random number that follows a truncated L\'evy distribution we generate L\'evy-stable random numbers and take the first one for which the condition $D_n < b |V|$ is fulfilled.

Using the numerically obtained eigenstates of the Hamiltonian \eqref{Ham_ExNN} we calculate the absorption strength according to eq.~\eqref{eq:A_j} and record the pair $(E_j, A_j)$ in a discrete histogram, with $E_j$ the respective eigenenergy. The underlying grid of the histogram is adapted  for each combination of $\StabIndex$, $\ScalePar$ and $N$
according to the following procedure:
The grid is centered around $E_{j=1}$ (Eq.~\eqref{eq:eigE}) and its width $\ell$ is determined using Eqs.~\eqref{eq:D_agg_weak_div_D_mon}, \eqref{eq:D_agg_intermediate_div_D_mon} and \eqref{eq:D_agg_strong_div_D_mon} such that
\begin{equation}
 \label{app:width}
 \ell = 20\,\ScalePar \times \left\lbrace\begin{array}{ll}
      \min\Big(1, \max\big(\ScalePar^{\frac{\StabIndex  - 1}{\StabIndex + 1}}, N^{\frac{1}{\alpha} - 1}\big)\Big) &,\; \textrm{for } \StabIndex \geq 1\\
      \max\Big(1, \min\big(\ScalePar^{\frac{\StabIndex  - 1}{\StabIndex + 1}}, N^{\frac{1}{\alpha} - 1}\big)\Big) &,\; \textrm{for } \StabIndex < 1
     \end{array} \right. .
\end{equation}
The grid boundaries are then taken as $E^{\rm abs}_{\rm min/max} = E_{j=1} \mp \ell/2$.
The factor $20$ in Eq.~\eqref{app:width} is a save margin to have the FWHM of the main absorption peak within the interval.
The resolution of the histogram, i.e.\ the width of the bins, is determined by $\ell/N_{\rm bin}$, with $N_{\rm bin}$ being the number of bins.
 For the presented data we have used $N_{\rm bin} = 10001$.

For the grid for the localization lengths we know that $N^{\rm loc}$ ranges from $0$ to $N$.
 For the data presented in this work we chose $100 \times N$ bins.

\paragraph{Extracting the FWHM:}

For the extraction of the FWHM from data containing a discrete grid $x$ and histogram $h(x)$, we first search for the grid value $x_{\rm max}$ for which the histogram $h(x)$ has its global maximum $h_{\rm max}$. Starting from $x_{\rm max}$ we search to the left (decreasing $x$ values) for the grid value $x_{\rm left}$ for which the histogram $h(x)$ fulfills for the first time $h(x=x_{\rm left}) \leq h_{\rm max}/2$. We repeat the search, but for increasing $x$ values to find $x_{\rm right}$, which also fulfills for the first time $h(x=x_{\rm right}) \leq h_{\rm max}/2$.
The difference $x_{\rm left} - x_{\rm right}$ gives the FWHM.
The application of this procedure to noisy data usually results in large uncertainties. Therefore, we used the following strategy: First we roughly estimate the FWHM, $\Dagg^{\rm rough}$, of the raw data, according to the above stated procedure. We then smooth the data by convolution with a Gaussian function with area one and standard deviation $\gamma = 1/120 \times \Dagg^{\rm rough}$. The prefactor $1/120$ is chosen large enough to smoothens the data considerably, but small enough to not alter the line-width significantly.
 Finally we extract the FWHM of the smoothened data.
To ensure, that the above described method does not produce wrong results for unexpected situations we visually compare the raw histogram with the smoothened histogram. We also visually check the extracted FWHM.

\section*{References}

\end{document}